\documentclass[10pt, sigconf]{acmart}
\usepackage{booktabs} % For formal tables
\setcopyright{rightsretained}
\usepackage{graphicx}
\usepackage{subfigure}
\usepackage{amsfonts}
\usepackage{amssymb}
\usepackage{color}
\usepackage{epstopdf}
\usepackage{amsmath}
\usepackage{balance}

\copyrightyear{2018}
\acmYear{2018}
\setcopyright{acmcopyright}
\acmConference[SenSys '18]{The 16th ACM Conference on Embedded
	Networked Sensor Systems}{November 4--7, 2018}{Shenzhen, China}
\acmBooktitle{The 16th ACM Conference on Embedded Networked Sensor
	Systems (SenSys '18), November 4--7, 2018, Shenzhen, China}
\acmPrice{15.00}
\acmDOI{00.0000/0000000.0000000}
\acmISBN{000-0-0000-0000-0/00/00}

\begin{document}
%----------------------------------------Make Title -----------------------------------------
%----------------------------------------------------------------------------------------------
\title{ShieldScatter: Improving IoT Security with Backscatter Assistance}
%---------------------------------------Make Abstract--------------------------------------

{\author{Zhiqing Luo}
\authornote{Co-primary authors.}	
\affiliation{
	\institution{Huazhong University of Science and Technology }
	\city{Wuhan}
	\country{China}}
	\email{zhiqing\_luo@hust.edu.cn}
	
\author{Wei Wang}
\authornotemark[1]
\authornote{This is the corresponding author of the work.}
\affiliation{
	\institution{Huazhong University of Science and Technology }
	\city{Wuhan}
	\country{China}}
\email{weiwangw@hust.edu.cn}
}

\author{Jun Qu}
\affiliation{
	\institution{Huazhong University of Science and Technology }
	\city{Wuhan}
	\country{China}}
\email{qjun@hust.edu.cn}

\author{Tao Jiang}
\affiliation{
	\institution{Huazhong University of Science and Technology }
	\city{Wuhan}
	\country{China}}
\email{taojiang@hust.edu.cn}

\author{Qian Zhang}
\affiliation{
	\institution{Hong Kong University of Science and Technology }
	\city{Hong Kong}
	\country{China}}
\email{qianzh@cse.ust.hk}
\renewcommand{\shortauthors}{Z. Luo et al.}
\begin{abstract}
%\justifying\let\raggedright\justifying    
The lightweight protocols and low-power radio technologies open up many opportunities to facilitate Internet-of-Things (IoT) into our daily life, while their minimalist design also makes IoT devices vulnerable to many active attacks due to the lack of sophisticated security protocols. Recent advances advocate the use of an antenna array to extract fine-grained physical-layer signatures to mitigate these active attacks. However, it adds burdens in terms of energy consumption and hardware cost that IoT devices cannot afford. To overcome this predicament, we present ShieldScatter, a lightweight system that attaches battery-free backscatter tags to single-antenna devices to shield the system from active attacks. The key insight of ShieldScatter is to intentionally create multi-path propagation signatures with the careful deployment of backscatter tags. These signatures can be used to construct a sensitive profile to identify the location of the signals' arrival, and thus detect the threat. We prototype ShieldScatter with USRPs and ambient backscatter tags to evaluate our system in various environments. The experimental results show that even when the attacker is located only 15 cm away from the legitimate device, ShieldScatter with merely three backscatter tags can mitigate 97\% of spoofing attack attempts while at the same time trigger false alarms on just 7\% of legitimate traffic.
\end{abstract}




\maketitle

%----------------------------------------Make Introduction-----------------------------------------
\section{Introduction}
\begin{figure}[t]
	\center
	\includegraphics[width=2.2in]{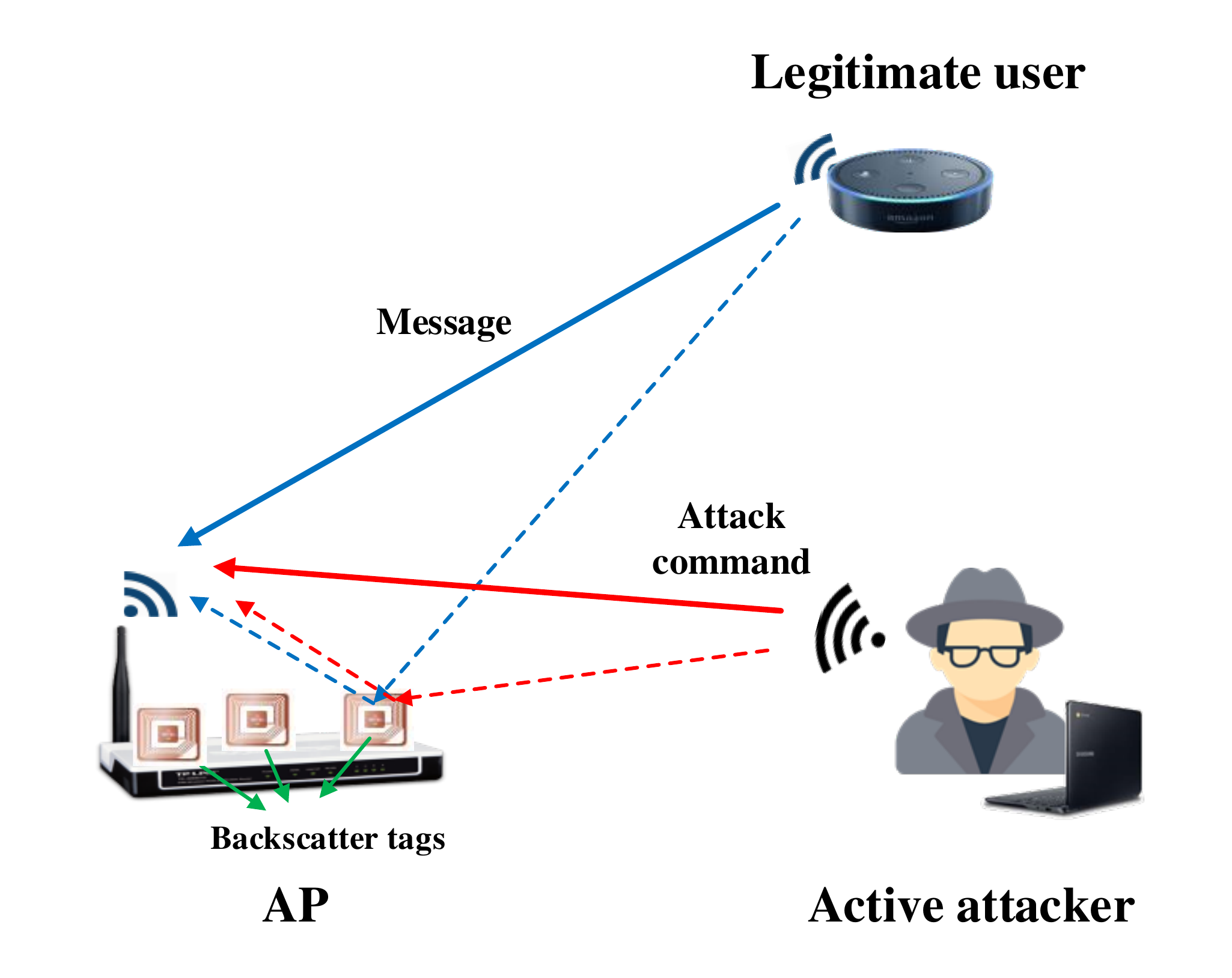} %\vspace{-0.2cm}
	\caption{Illustration of active attack.}
	\label{fig:sketch} \vspace{-0.3cm}
\end{figure}

The continuous advancement in low power radios and lightweight protocols is driving the proliferation of Internet-of-Things (IoT) in our daily life. However, the other side of the coin is that IoT devices easily bear the risks by active attacks such as spoofing attack and Denial-of-Service (DoS) attack during devices pairing or data transmission. For example, considering the scenario shown in Figure~\ref{fig:sketch}, a legitimate user~(e.g., a smart TV) is pairing or sharing data with an IoT access point~(AP). An active attacker equipped with an omnidirectional or directional antenna impersonates the legitimate user and sends a fake command (e.g., DoS command or fake data) to the AP.

%Figure 1: AP and devices communication but attacked by an active attacker，

Traditional approaches mainly rely on the complex encryption algorithm, which will lead to computational resources and energy waste~\cite{gehrmann2004manual,dietrich2007financial} and are not feasible for simply designed IoT devices. Alternatively, fine-grained physical-layer signatures, such as angle of arrival~(AoA)~\cite{xiong2013securearray}, channel state information CSI ~\cite{jiang2013rejecting} and received signal strength (RSS) ~\cite{cai2011good,chandrasekaran2009detecting} have recently received much attention to mitigating these threats. However, these systems require at least two or an antenna array (e.g., an eight-antenna array) to construct sensitive signatures, and thus are expensive and not applicable for the systems where the APs and IoT devices are equipped with only a small number of antennas. In addition, in an open space~(e.g., the hall), the multi-path phenomenon is indistinctive, making it difficult to extract the fine-grained AoA signatures. 

This paper presents ShieldScatter, a lightweight system to secure IoT device pairing and data transmission. Instead of relying on the expensive antenna array, ShieldScatter advocates the use of merely several ultra-low-cost and battery-free backscatter tags~\cite{liu2013ambient} to secure the IoT device. Our key insight is that backscatter tags communicating based on backscattering ambient signals, can be exploited to intentionally create fine-grained multi-path signatures. In particular, upon detecting any suspicious transmission, the legitimate user is asked to transmit the challenge-response based signals to the AP within the coherence time. At the same time, the AP controls the tags to backscatter the signals. We observe that even in dynamic channel environments, the propagation signatures created by the backscatter tags can be used to construct a unique profile to identify each user. This unique propagation signature profile then confirms whether these two signals are from the same devices and help defend against active attacks.

To realize the above idea, we entail the following challenges.

\textit{(1) How to employ backscatter tags to create sensitive propagation signatures without using an expensive antenna array or other powerful hardware?}
Most home devices employ only a small number of antennas for data transmission, which makes it inapplicable for them to construct accurate propagation signatures, such as AoA. To overcome this predicament, we consider leveraging the multi-path features of backscatter tags to generate distinct propagation signatures. In particular, we attach the tags around the AP. When initializing the system, the AP receives the signals. At the same time, the AP controls the tags to reflect wireless signals in turn. This intentional deployment can create artificial multi-path propagations that are sensitive to senders' locations. If these two signals are from the same devices, the multi-path effects of the backscatter tags on these two signals will have strong similarity in the coherence time. Thus, these distinct propagation signatures can be used to identify the legitimate users.

\textit{(2) How to construct reliable signatures when unstable factors exist?}
In home environments, walking people, environmental noise and an imperfect circuit design of the tags will lower the similarity of these two signals from the legitimate users. In order to construct reliable signatures, ShieldScatter extracts the representative features from the signals for the first step. Then, ShieldScatter aligns and compares the similarity of the features using dynamic time warping~(DTW). Furthermore, a one-class support vector machine~(SVM) classifier is used to distinguish and defend against signals from active attackers. If the signals are from the same devices, it will lead to strong similarity and short DTW distances, and then these signals will be clustered into the legitimate class. Otherwise, it will lead to large DTW distances and ShieldScatter can detect and defend against the attackers.

\textbf{Summary of result.}
We prototype ShieldScatter with USRPs and ambient backscatter tags to evaluate our system in various environments. The experimental results show that even when the attacker is located only 15 cm away from the legitimate device, ShieldScatter with merely three backscatter tags can mitigate 97\% of spoofing attack attempts while at the same time trigger false alarms on just 7\% of legitimate traffic.

\textbf{Contributions.}
First, we propose ShieldScatter to use the multi-path propagation signatures intentionally created by backscatter tags to secure the IoT devices. Second, we use multiple backscatter tags to create  unique multi-path signatures, which avoids the employment of an expensive antenna array to obtain fine-grained signatures and can work in the absence of multipath. Finally, our results show that our system is robust even when active attacker is close to the legitimate user. 

\section{Motivation}
In this section, we first discuss the potential threats to the IoT devices and argue that a lightweight mechanism designed for securing these devices is critical. Next, we investigate the fine-grained signatures of physical-layer radio propagations used to defend against active attacks, which motivates our design of using backscatter tags to secure IoT devices. 

\subsection{Threat Model}\label{sec:threat}
\begin{figure}[t]
	\center
	\includegraphics[width=1.8in]{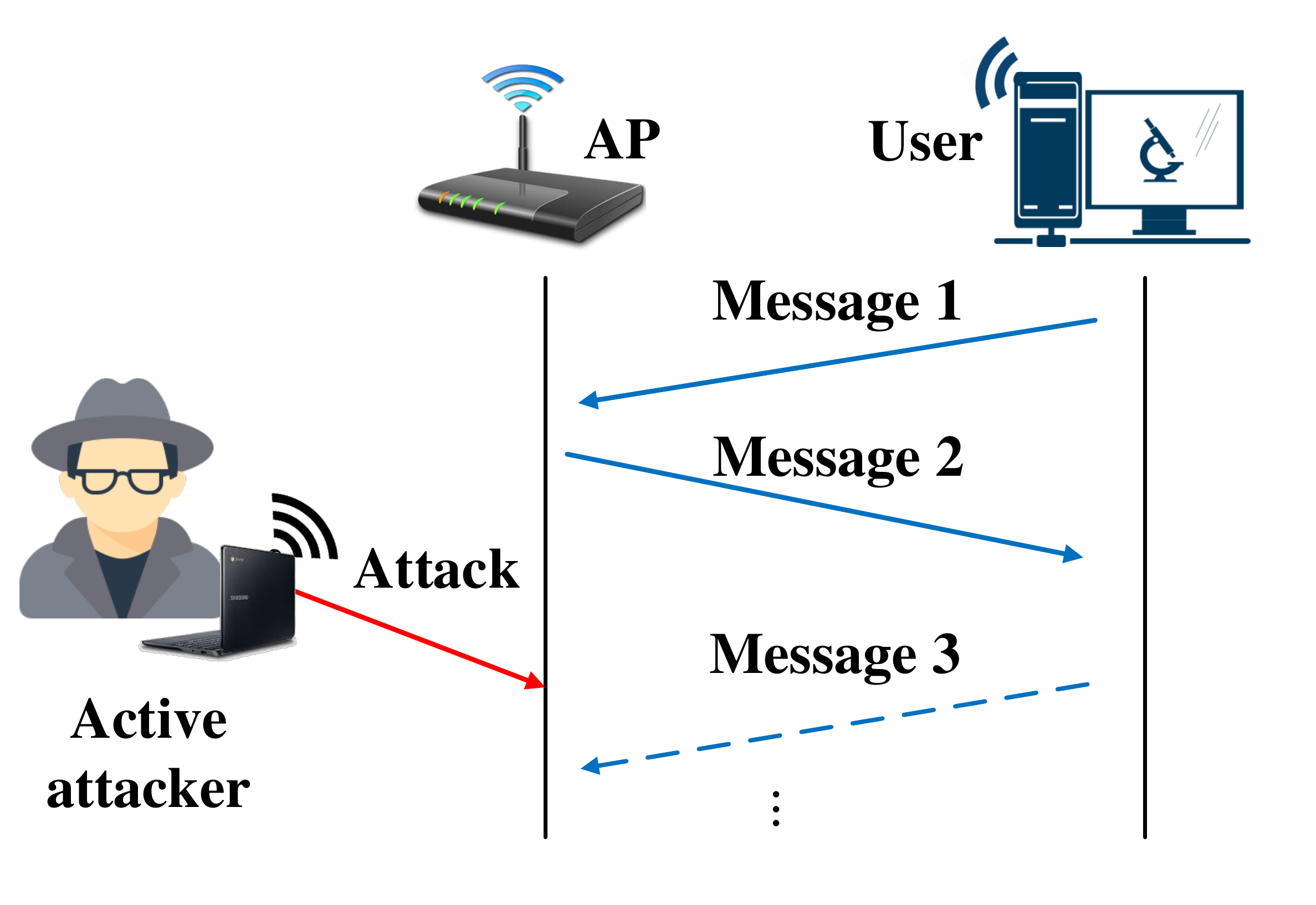} %\vspace{-0.2cm}
	\caption{The active attacker sends DoS command or fake data to the AP when the IoT device is pairing or exchanging data with the AP. }
	\label{fig:attack_model}\vspace{-0.3cm}
\end{figure}

Recently, the link establishment and data sharing between smart home IoT devices have become more and more indispensable. However, they are easily attacked by the active attackers. For example, as shown in Figure~\ref{fig:attack_model}, the legitimate user is pairing with an AP based on the challenge-response protocols. When initializing pairing, the legitimate user who intends to share information with an AP sends a message~(\textit{Message 1}) for request. If the AP receives this message, it sends an acknowledgement message~(\textit{Message 2}) back. Finally, the legitimate user receives \textit{Message 2} and feeds back the message~(\textit{Message 3}). However, during this process, a powerful attacker who has all the priori knowledge of the protocol and the legitimate user~(e.g., the coding scheme, carrier frequency and signal strength) can use an omnidirectional antennas to detect the initialization of the communication. Then,  it enables a directional antenna to inject fake data~(e.g., the DoS command or spoofing data) to attack the AP, and thus the AP receives the command from the attacker, %instead of the message (e.g., \textit{Message 3}) from the legitimate user
which leads to the rejection or unauthorized access to the AP.

ShieldScatter considers that the attacker will not be triggered to attack the IoT devices all the time. In other words, only if the attacker detects a legitimate user trying to connect or share the data with the AP~(e.g., detecting \textit{Message 1}), then the attacker will initialize the attack. Besides, ShieldScatter only defends against active attacks and makes no exploration of protecting against passive attacks such as eavesdropping attacks and information leakage.

\subsection{ Propagation Signatures }\label{sec:signature}
\begin{figure}[t]
	%    \center 
	\subfigure[AoA signatures.]
	{\includegraphics[width=1.6in]{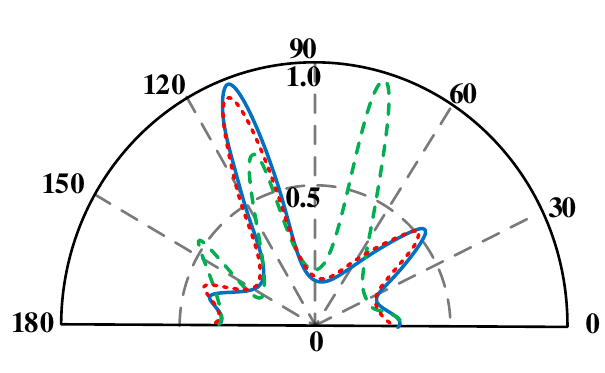}}
	\subfigure[Backscatter signal energy.]
	{\includegraphics[width=1.6in]{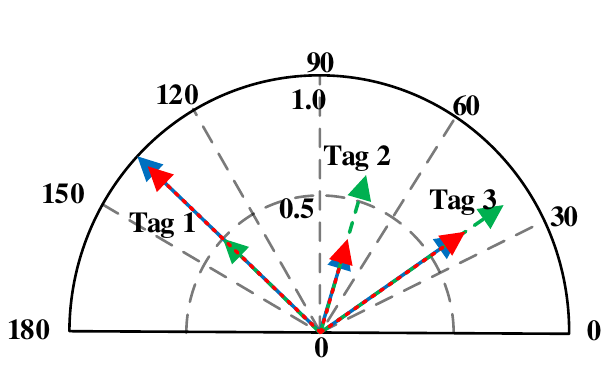}}
	\caption{If the signals are from the same devices, the AoA signatures have high similarity between them. Otherwise, the AoA signatures will be different. When we use three tags to create multipath, the average energy of each tags has the similar performance as AoA signatures.}
	\label{fig:AoA}\vspace{-0.3cm}
\end{figure}

Existing approaches to secure these active attacks by relying on extracting fine-grained propagation signatures from the physical-layer information. In particular, as shown in Figure~\ref{fig:AoA}(a), SecureArray~\cite{xiong2013securearray} extracts the AoA signatures from the received signal with an antennas array. If the received signals are from the same location within the coherence time, the radio propagation will experience the same multipath, which accordingly leads to strong similarity for the AoA signatures (e.g., the red and blue dashed). Then, based on the similarity of the AoA signatures,  active attacks can be detected. 

However, in smart homes, the devices may lack multiple antennas, and thus the methods of extracting fine-grained signatures with an antenna array will be inapplicable. Inspired by SecureArray that uses multi-path signatures to secure the devices, we observe that low-cost and battery-free backscatter tags can also generate such multi-path propagation signatures without using an expensive antenna array. According to~\cite{liu2013ambient}, as shown in Figure~\ref{fig:sketch}, backscatter is a new communication primitive where a tag transmits data by intermittently reflecting ambient signals. Accordingly, the multipath created by the tags can be used to construct a sensitive profile to protect the IoT devices.

In order to verify this idea, we employ three tags to attach around the AP at a distance of a half-wavelength and use two USRPs to emulate legitimate user and active attackers, respectively. Then, the AP controls the tags to reflect the signals in turn. Finally, we extract backscatter signals using the sliding windows and compare the average energy of each tag. The result is shown in Figure~\ref{fig:AoA}(b). We observe if the signals are from the same device, the average energy of the tags will have high similarity (e.g., the red and blue dashed). Whereas, if the signals are from different devices, even though the attacker is in the same direction from the AP, the distances from the tags to the user are not the same as the distances from the tags to the attacker, and thus it will lead to significant differences with respect to the amplitudes of the tags (e.g., the green dashed). This result presents a similar performance to AoA signatures and it motivates us to design ShieldScatter, a lightweight system to secure the IoT devices by using backscatter tags. ShieldScatter intentionally creates multi-path signatures by controlling the tags to reflect the ambient signals in turn. Besides, in order to construct a reliable profile to detect the threat, ShieldScatter also fetches other representative features and combines with a one-class SVM based classifier to identify  the attackers.

Besides, we should notice that the signal transmission should be completed within the coherence time. This is because the wireless channel is easily affected by the environmental states. However, when in coherence time, the channel can be treated as stable even in dynamic environments. The coherence time can be defined as, $T={{9\lambda}\over{16\times\pi\times{{v}}}}$, where $ \lambda$ represents the carrier wavelength and $v$ indicates the maximum velocity of legitimate user~\cite{steele1999mobile,xiong2013securearray}.
\begin{figure}[t]
	\center
	\includegraphics[width=3.5in]{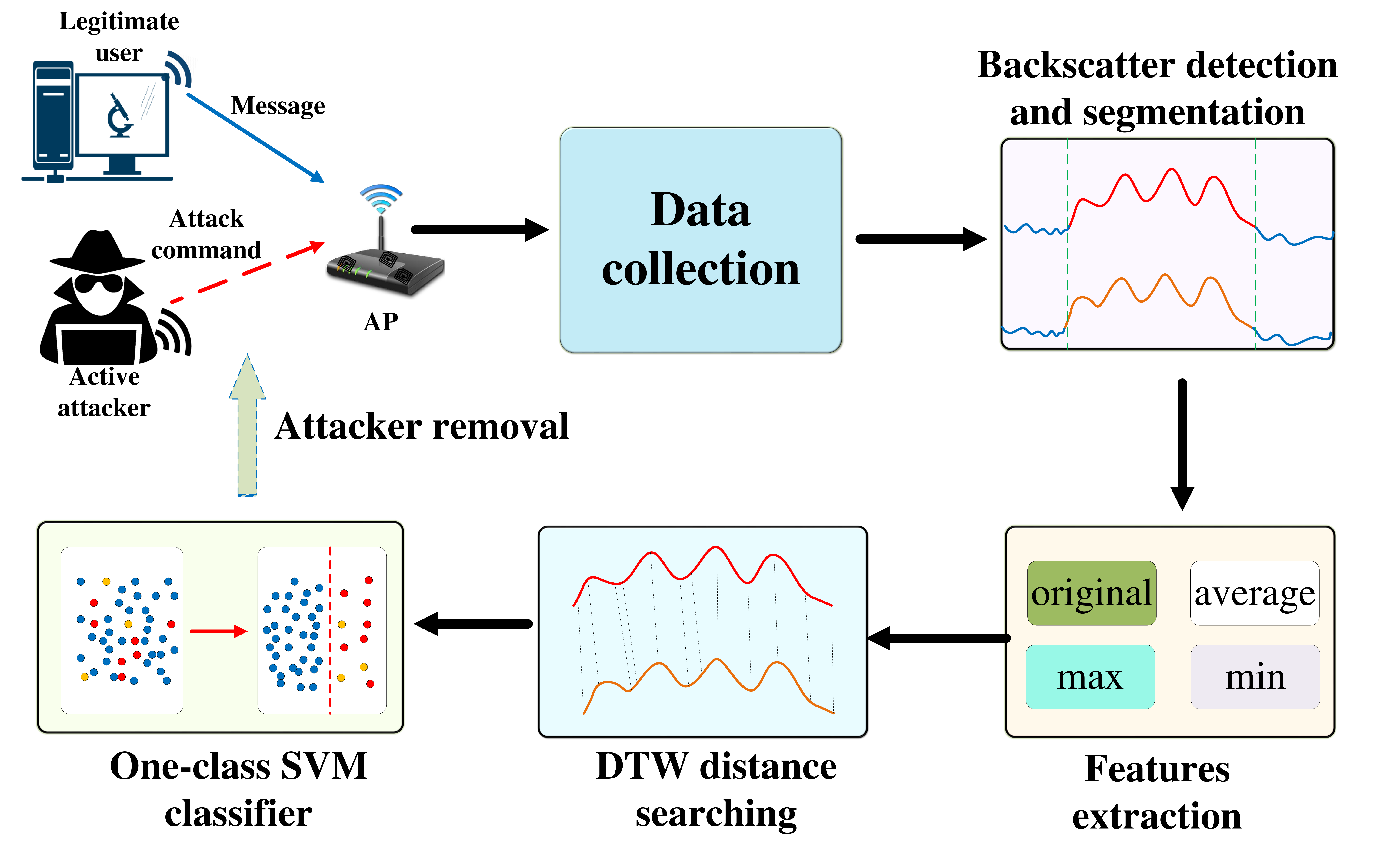} %\vspace{-0.2cm}
	\caption{System overview.}
	\label{fig:sysetm}\vspace{-0.3cm}
\end{figure}

% Figure 4  system framework
\section{System Design}\label{sec:system}
In this section, we first present an overview of ShieldScatter which consists of four key steps. Then, we elaborate on each step and provide the technical details in the following subsections.
\subsection{ System Overview }\label{sec:overview}
The basic idea of ShieldScatter is to construct sensitive multi-path propagation signatures using several backscatter tags attached around the AP instead of an expensive antenna array. In particular, as shown in Figure~\ref{fig:attack_model}, when the legitimate user is pairing with the AP, it is easily attacked by the fake transmission (e.g., by launching a deauthentication message). To defend against this attack, upon detecting the suspicious transmission, the AP is asked to control the tags to work in turn during this processing. Then, by comparing the signatures from the same device (e,g.,  \textit{Message 1} and \textit{Message 3}) or from different devices (e,g., \textit{Message 3} and the suspicious command), ShieldScatter carries out the security system to verify whether the suspicious messages are from the legitimate device and then defend against active attacks, which will be elaborated in Section~\ref{sec:security}.

To detect the attacks, as illustrated in Figure~\ref{fig:sysetm}, at a high level ShieldScatter needs to go through the following four steps. First, based on the collected data at the AP, ShieldScatter detects and segments the signals that includes the backscatter signal. Second, ShieldScatter extracts representative features from the segments. Third, to construct a reliable propagation profile, ShieldScatter compares the features by computing the distances with DTW. Finally, based on the profiles with respect to the DTW distances, ShieldScatter can identify and defend against the attacks with a one-class SVM classifier.

\subsection{ Backscatter Detection and Segmentation }\label{sec:segmentation}
Recall that ShieldScatter constructs sensitive multi-path propagation signatures by controlling the backscatter tags to work in turn. However, because of the latency of the tag initialization, the received signals always contain the component without backscatter, which makes the multi-path signatures not distinct in this component. Thus, to ensure to extract reliable feature from the received signals, ShieldScatter needs to detect and segments the received signals for the first step.

\begin{figure}[t]   
	%    \center 
	\subfigure[Backscatter decoding.]
	{\includegraphics[width=2.7in]{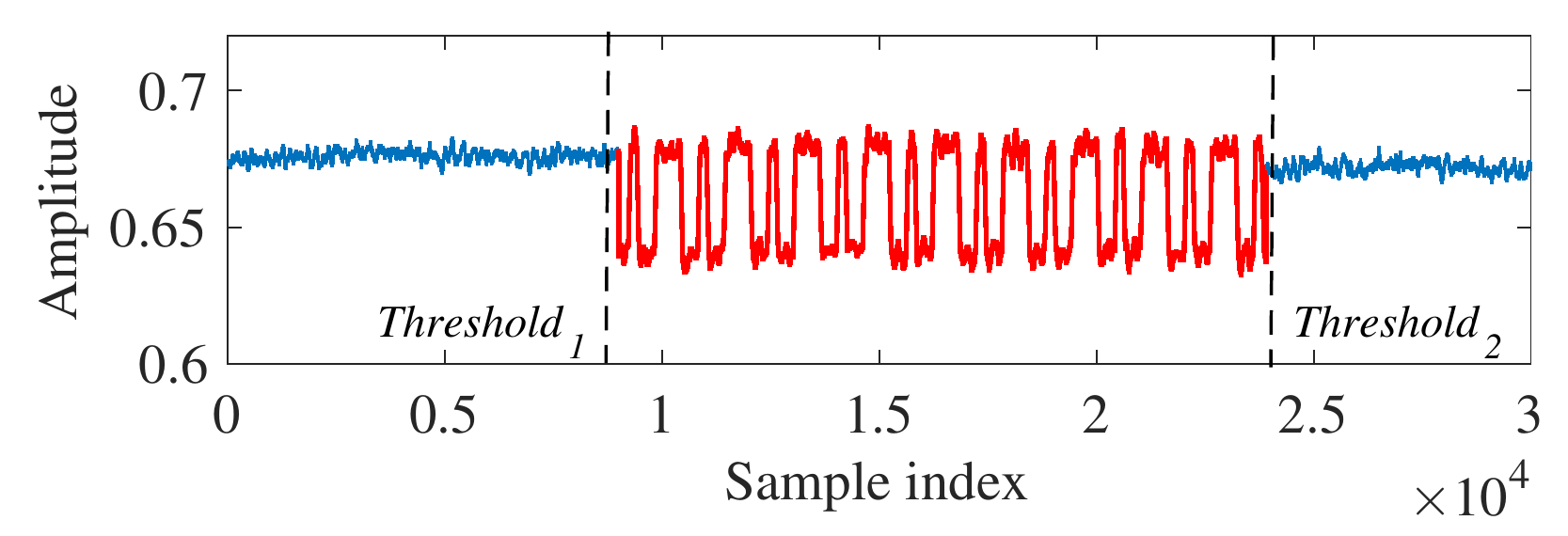}}
	\subfigure[Energy envelope detection.]
	{\includegraphics[width=2.7in]{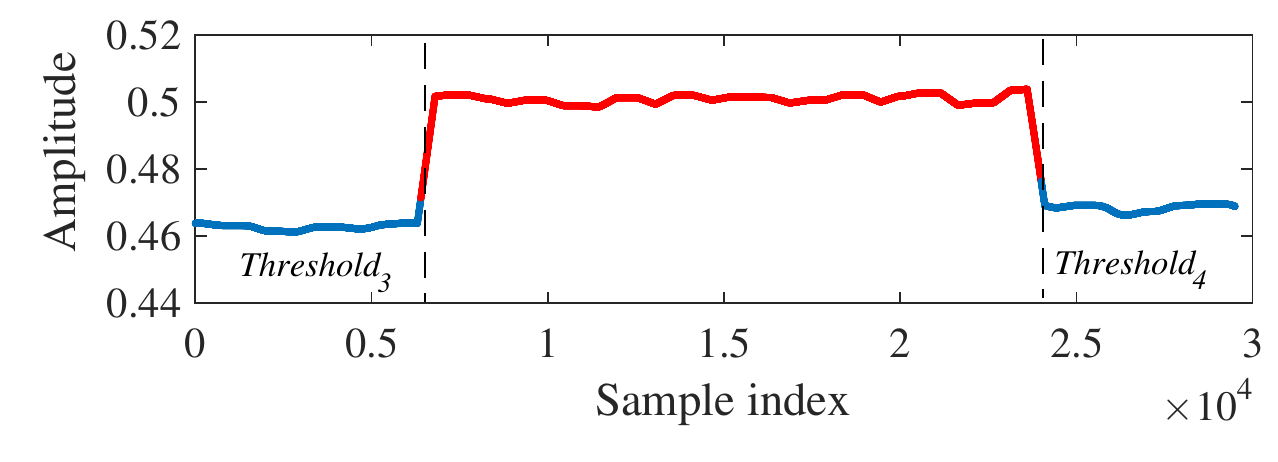}}
	\caption{ShieldScatter detects and segments backscatter component by combining backscatter decoding and energy envelope detection.}
	\label{fig:segment}\vspace{-0.3cm}
\end{figure}

\begin{figure*}[t]
	%    \center 
	%\centering  
	\subfigure[Original signals.]
	{ \includegraphics[width=5.0cm]{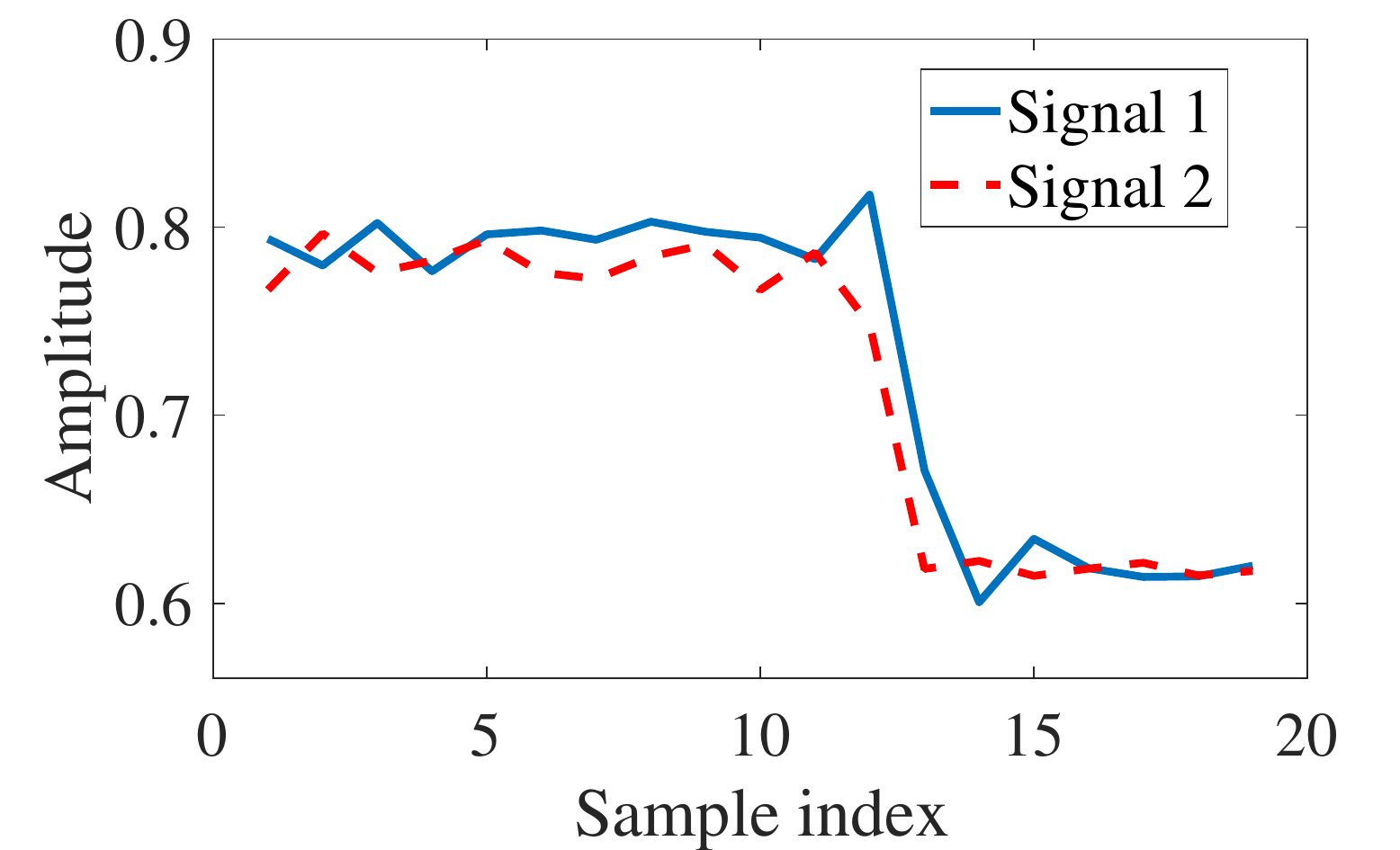}  }
	\subfigure[Smoothing signals.]
	{ \includegraphics[width=5.0cm]{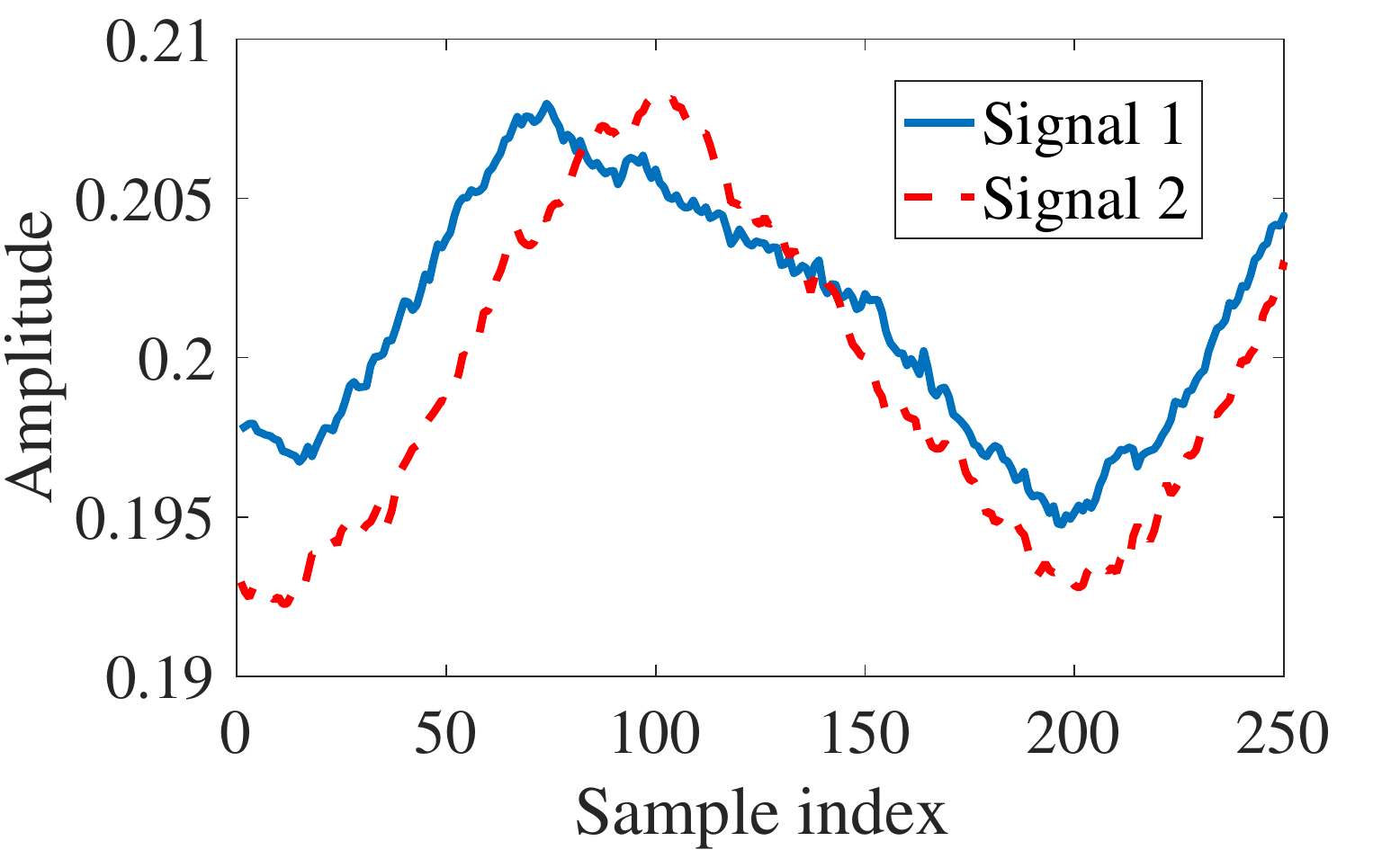}  }
	\subfigure[Maximum.]
	{ \includegraphics[width=5.0cm]{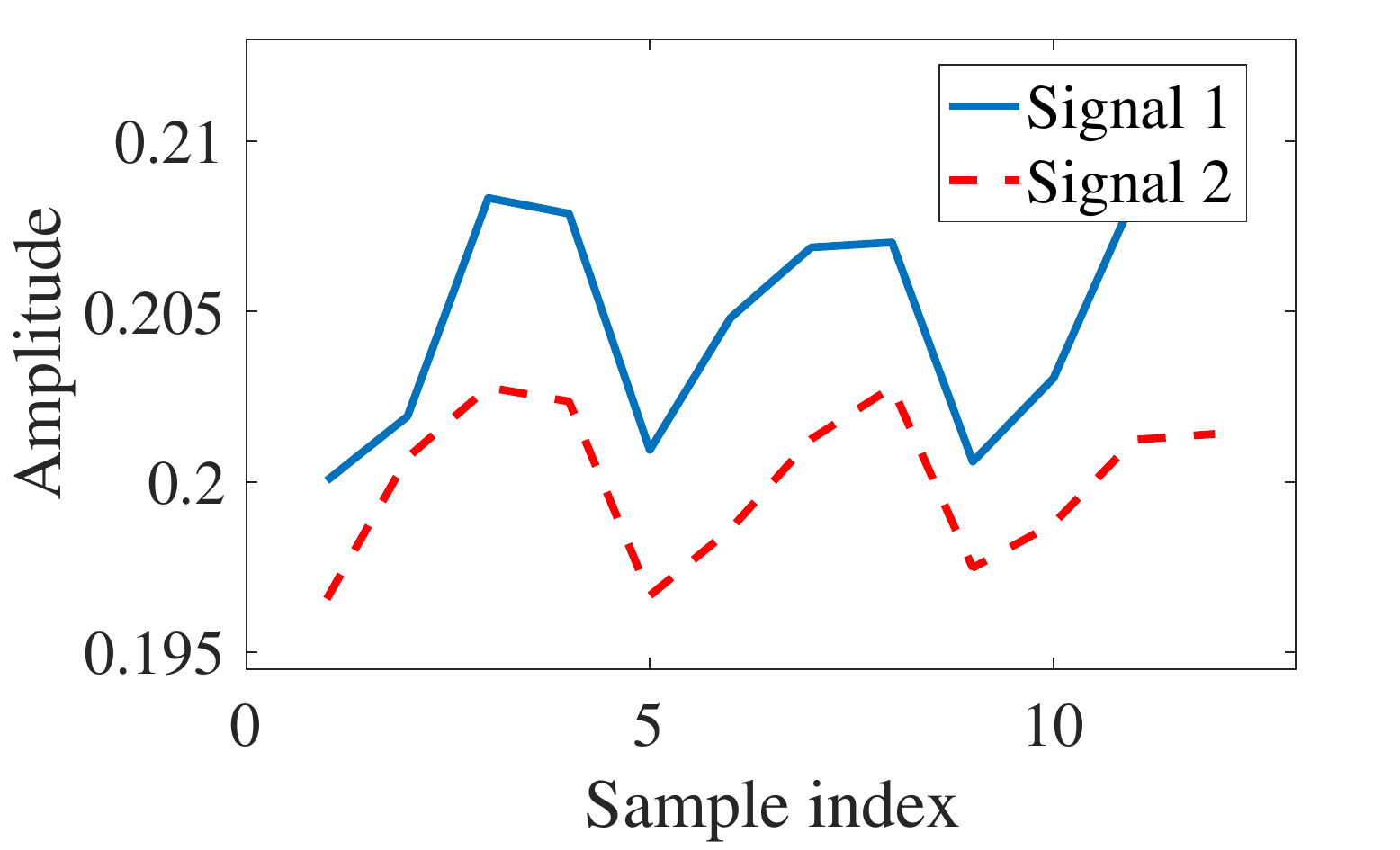}  }
	\caption{Features extracted from the same device.}
	\label{fig:feature_positive}\vspace{-0.3cm}
\end{figure*}
\begin{figure*}[t]
	%    \center 
	%\centering  
	\subfigure[Original signals.]
	{ \includegraphics[width=5.0cm]{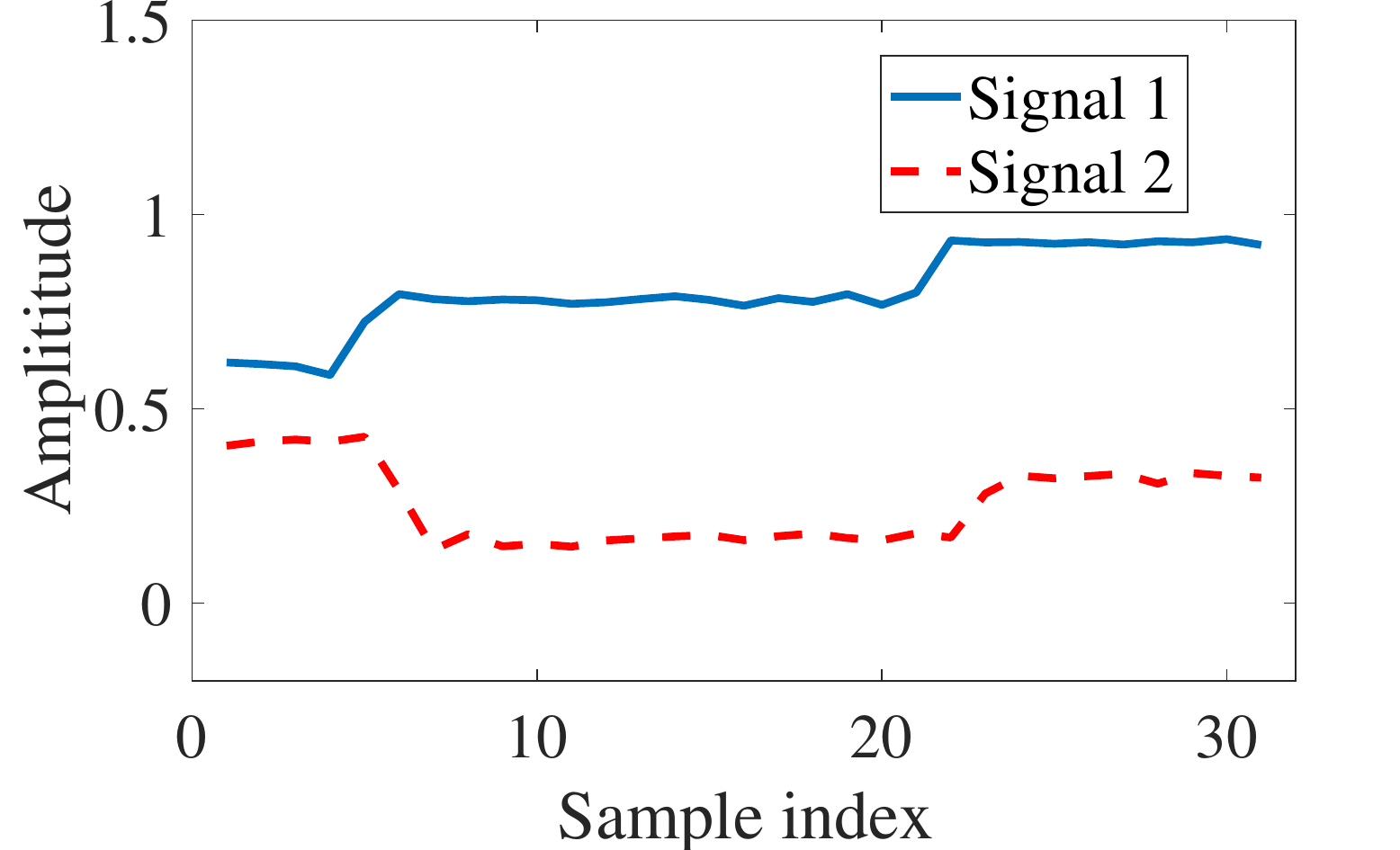}  }
	\subfigure[Smoothing signals.]
	{ \includegraphics[width=5.0cm]{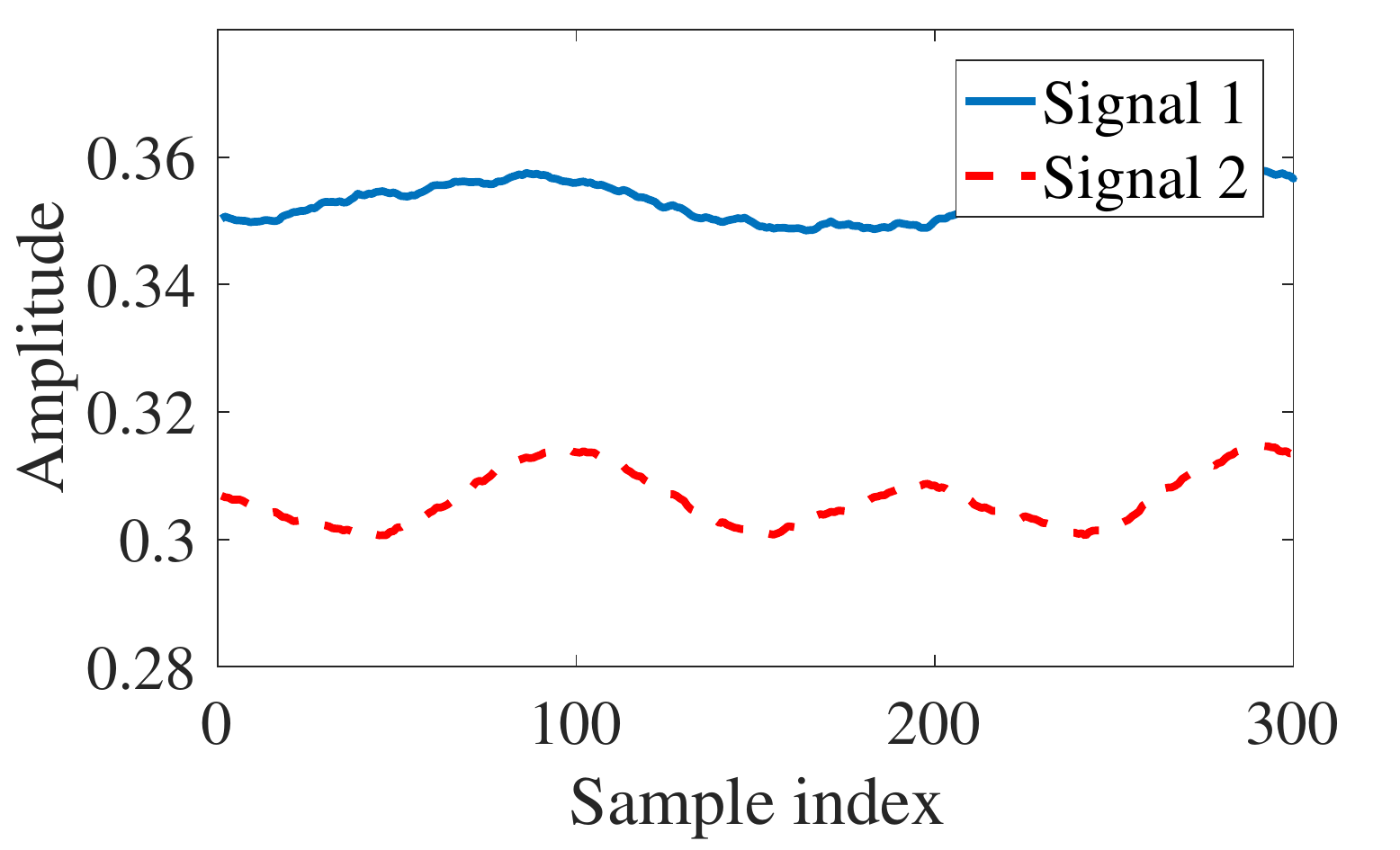}  }
	\subfigure[Maximum.]
	{ \includegraphics[width=5.0cm]{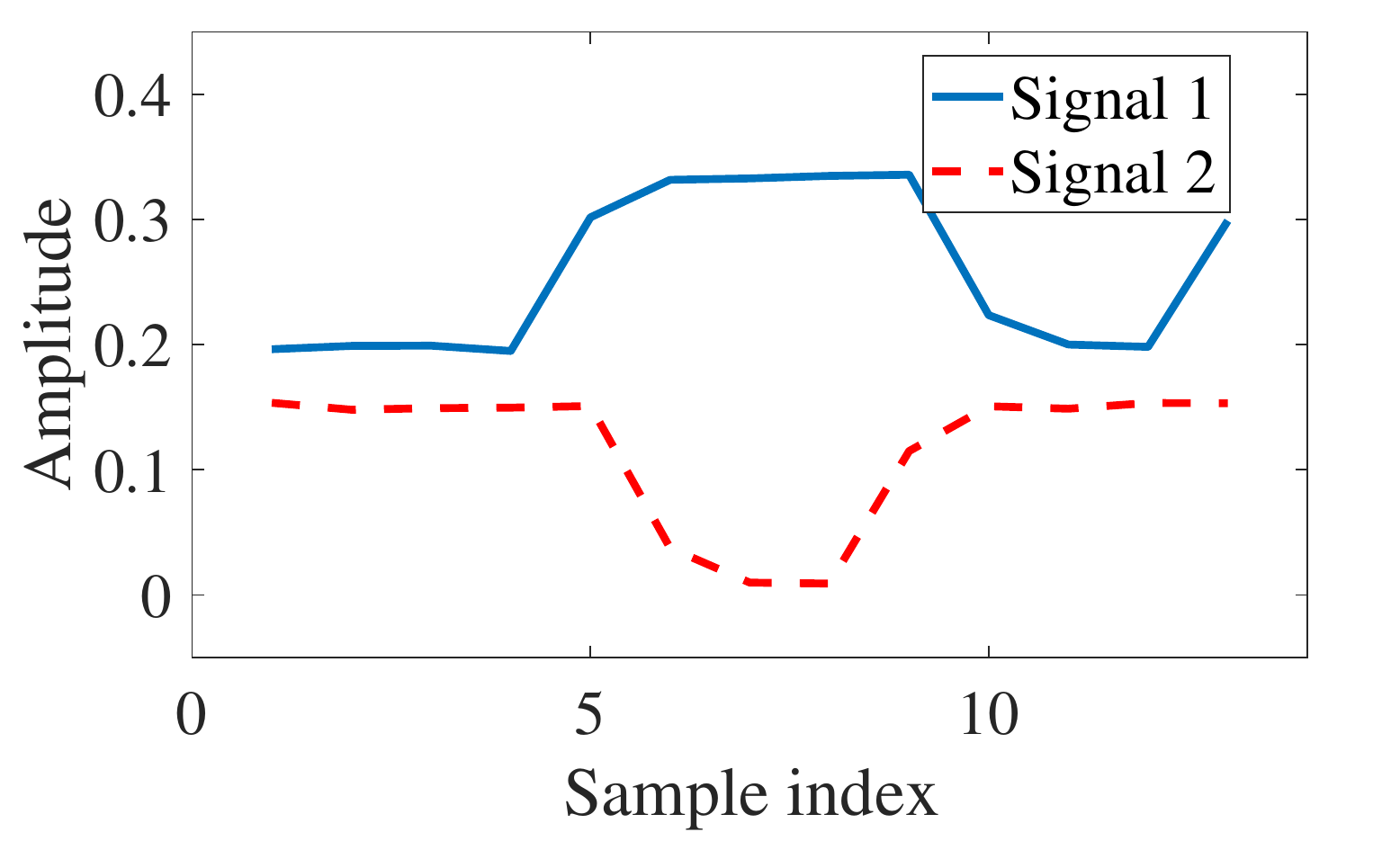}  }
	\caption{Features extracted from the different devices.}
	\label{fig:feature_negative}\vspace{-0.3cm}
\end{figure*}
In our system, in order to segment the signals, we first decode the received signals using a moving average method as~\cite{liu2013ambient} where we use a sliding window with a length of 50 samples to smooth the signal. After smoothing, ShieldScatter decodes the message of backscatter signals as shown in Figure~\ref{fig:segment}(a). In order to determine the segment, ShieldScatter follows the principle: if the AP can continuously decode the backscatter signals, then the corresponding original signals samples from the starting point to the ending point are considered as the segment that contains the backscatter signal. Accordingly, we mark the starting point and the ending point as $\eta_1$ and $\eta_2$, respectively.

After that, ShieldScatter can achieve a raw signal segmentation to detect the backscatter. However, because of the imperfect circuit design and noise, it is not accurate enough to segment the signals. Thus, to improve the accuracy of the signal segmentation, we employ an energy envelope detecting method for assistance. Specifically, a sliding window upon the received signal amplitude is used to detect the backscatter, where we calculate the average energy $E(i)$ within this sliding window by
\begin{equation}
\\E(i)={{1\over N}{ \sum_{i=1}^{i+N}{|x(i)|^2}}},
\end{equation}
where $N$ is the length of the sliding window, and $x(i)$ is the amplitude of the sample at sample index $i$. After calculating the energy of the signals, we can easily yield the energy envelope as shown in Figure~\ref{fig:segment}(b). It is obvious that the energy envelope changes greatly when the backscatter tags are working. Besides, we also find that the backscatter signal always locates at the center of samples without backscatter, which inspires us that the energy envelope will experience a large variance when backscatter is working.  Thus, in order to determine the starting point to ending point of the segment that contains the backscatter signal, we calculate the variance of the energy envelope by
\begin{equation}
\\V(j)=Var[E(j):E(j+N) ],
\end{equation}
where $Var[E(j):E(j+N) ]$ represents we calculate the variance in every $N$ samples and $ V(j)$ represents the variance at index $j$. Then,  we determine the starting point and ending point of signals containing backscatter following the constraint
%\begin{gather}
%0<j<\eta_3, V(j)<e^2,V(\eta_3+1)>e^2 \notag  , \\
%\eta_4<j<end, V(j)>e^2,V(\eta_4+1)<e^2
%\end{gather}
\begin{eqnarray}
& & 0<j<\eta_3, V(j)<t,V(\eta_3+1)>t, \notag \\
& & \eta_4<j<m, V(j)>t,V(\eta_4+1)<t
\label{eq:diaggonaldominance}
\end{eqnarray}
where $m$ represent the total number of $V(j)$, $\eta_3$ and $\eta_4$ represents the starting point and the ending point of the segment, respectively. $t$ is the dynamic threshold. According to our experimental study, we set the threshold as $e^2$ where $e$ is the minimum energy of all the tags and it can be obtained by the backscatter decoding.

Finally, we combine the method of backscatter decoding and energy envelope detecting to determine the segment by
\begin{gather}
\eta_s={\left.(\eta_1+\eta_3)\right/2}\notag ,\\
\eta_e={\left.(\eta_2+\eta_4)\right/2} ,
\end{gather}
where $\eta_s$ and $\eta_e$ represent the final decision for the starting point and the ending point of the segment, respectively. 

\subsection{Feature Extraction}\label{sec:features}
Before constructing reliable multi-path propagation profiles, ShieldScatter should fetch representative features from the segments. According to the ambient backscatter theories [12], the backscatter signal is an additional multipath generated by the tag. This additional multipath can either constructively or destructively interfere with the ambient signal. All of these multipath mainly affect the amplitude of the ambient signal. Thus, to construct reliable signatures, from the obtained segments, ShieldScatter selects the features with respect to the signal amplitude.

Intuitively, ShieldScatter can directly use the received raw data (we call it as original signals in this paper) for comparison. However, the existing ambient noise will lower the similarity of the comparison, and thus only the original signal is not enough to construct a reliable propagation profile. Thus, in our system, besides \textit{original signal}, five other significant features with respect to the signal amplitude, including \textit{smoothing signal}, \textit{energy envelope}, \textit{variance} of the signals, \textit{maximum} and \textit{minimum}, are also extracted to construct our unique and sensitive multi-path propagation signatures to profile each legitimate user. Specifically, in order to acquire smoothing signal, ShieldScatter filters the original signals by using a sliding window. As for energy envelope, variance, maximum and minimum, ShieldScatter extracts these features by computing the average energy envelope, variance, maximum, and the minimum in every 50 data samples of the original signals. Accordingly, ShieldScatter can obtain six \textit{feature series} for each of the segments obtained in Section~\ref{sec:segmentation}. As shown in Figure~\ref{fig:feature_positive} and Figure~\ref{fig:feature_negative}, the feature series are extracted from the signals of the same and different devices within the coherence time, respectively. It is obvious that the features are similar when the signals are from the same devices. Otherwise, if the signals are from different devices, the features extracted from these two signals are quite different. Accordingly, these representative features can be used to construct reliable profile and secure the legitimate user.

\subsection{DTW Distance Searching}\label{sec:DTW}
After acquiring the feature series from the segments, ShieldScatter needs to compare the features to detect active attackers. Thus, a reliable method to evaluate the similarity of the extracted features of the corresponding segments is needed. Intuitively, a simplest method is to calculate the correlation of every two corresponding feature series directly. However, because of the noise and imperfect circuit design of the backscatter tags, even though the methods combining the decoded and energy envelope have been exploited to detect and segment the signal, they still cannot guarantee an absolutely accurate partition of the received signal. Besides, since the transmitting signals are sinusoidal waves and our tags reflect the signal in a periodic way, the imperfect segmentation of the signals leads to the shifting of the features, which can be seen in Figure~\ref{fig:feature_positive}. Thus, simply computing the correlation to compare the feature series will lower the similarity and are not applicable to our system. Instead, a method that can mitigate the unfavorable effects caused by misalignment is more desirable for comparison in our design.
\begin{figure}[t]
	\center
	\includegraphics[width=2.5in]{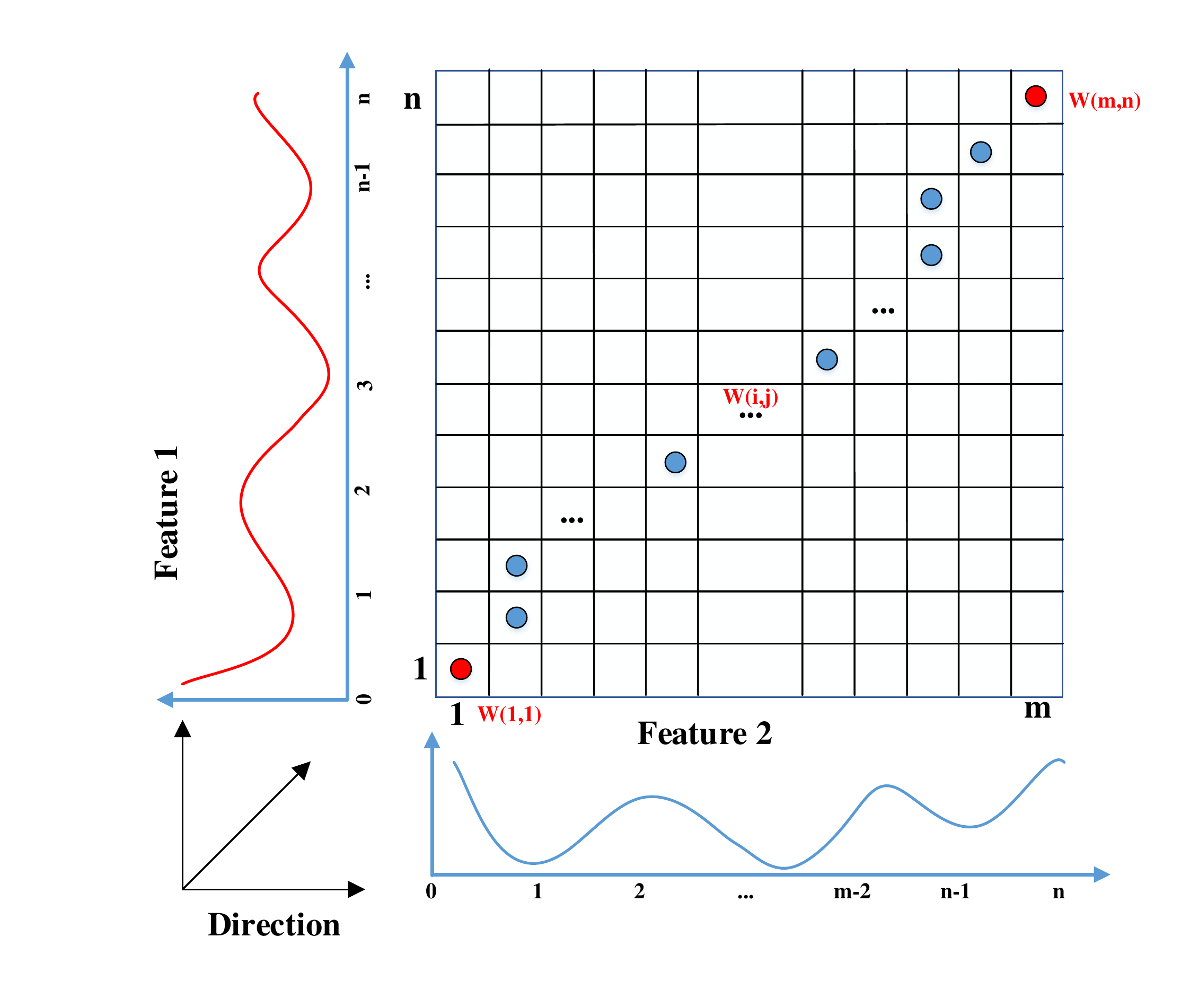} %\vspace{-0.2cm}
	\caption{ShieldScatter searches the shortest distance to compare the similarity between the extracted features with DTW.}
	\label{fig:DTW}
	%\vspace{-0.3cm}
\end{figure}
Inspired by PinIt~\cite{wang2013dude} and the method for the word matching in speech recognition, they have similar nature of signatures shifting. In order to mitigate the effect of misalignment, a most commonly used method DTW can be adopted to overcome this predicament. Thus, we compare the similarity of the features to construct the propagation profiles by using DTW distance computing~\cite{salvador2007toward} as follows:
supposed given the two feature series $X(i)$ and $Y(j)$ of the corresponding segments (e.g., the feature \textit{maximum} of \textit{Message 3} and the suspicious message in Figure~\ref{fig:attack_model}), the goal of DTW is to find the minimum cost of the mapping sum from the feature series $X(i)$ to $Y(j)$. In particular, the cost of DTW from each sample of $X(i)$ to the arbitrary sample in $Y(j)$ is defined by using the Euclidean distance
\begin{equation}
\\w(i,j)=|X(i)-Y(j)|.
\end{equation}
Then, based on the Euclidean distance between each two samples of these two feature series, a dynamic programming algorithm is used for DTW to search for the warp path distance. To understand the shortest path searching in ShieldScatter, we define the two feature series $X(i)$ and $Y(j)$ of the features
\begin{equation}
\\X(i)=X(1), X(2), ...X(i)... X(m),
\end{equation}
\begin{equation}
\\Y(i)=Y(1), Y(2), ...Y(i)... Y(n),
\end{equation}
where $m$ and $n$ represent the length of the these two series, respectively. Then, based on these two series, we construct a network matrix $W$ as shown in Figure~\ref{fig:DTW}, where each matrix $(i,j)$ in $W$ indicates the Euclidean distance $w(i,j)$ corresponding to $X(i)$ and $Y(j)$. In order to search the best way to align $X(i)$ and $Y(j)$, DTW first starts from the point
$w(1,1)$. Then, DTW searches the shortest way following these rules: (1) the next step in any matrix should be $right$, $up$ or $right$ $ and $ $up$, which guarantees the monotonicity of the constraint for DTW; (2) the total cost of the distances should be the lowest. We define DTW in mathematical expressions as 

\begin{eqnarray}
& \underset{W}{\text{min}} & \sum_{i=1}^{m} \sum_{i=1}^{n}   w(i,j) \\
& \text{s.t.} & sp=w(1,1),{ep}=w(m,n), \notag \\
& & {st}(i)\le st({i+1}),st(j)\le{st}({j+1}).
\end{eqnarray}
where $W$ represents the route matrix, $sp$ and $ep$ the starting-point and ending-point, respectively. $st(i)$ indicates the horizontal axis coordinates at the $i_{th}$ step, and the two constraint conditions have guaranteed the boundary and monotonicity for the route selection in DTW.

\begin{figure}[t]
	\center
	\includegraphics[width=2.6in]{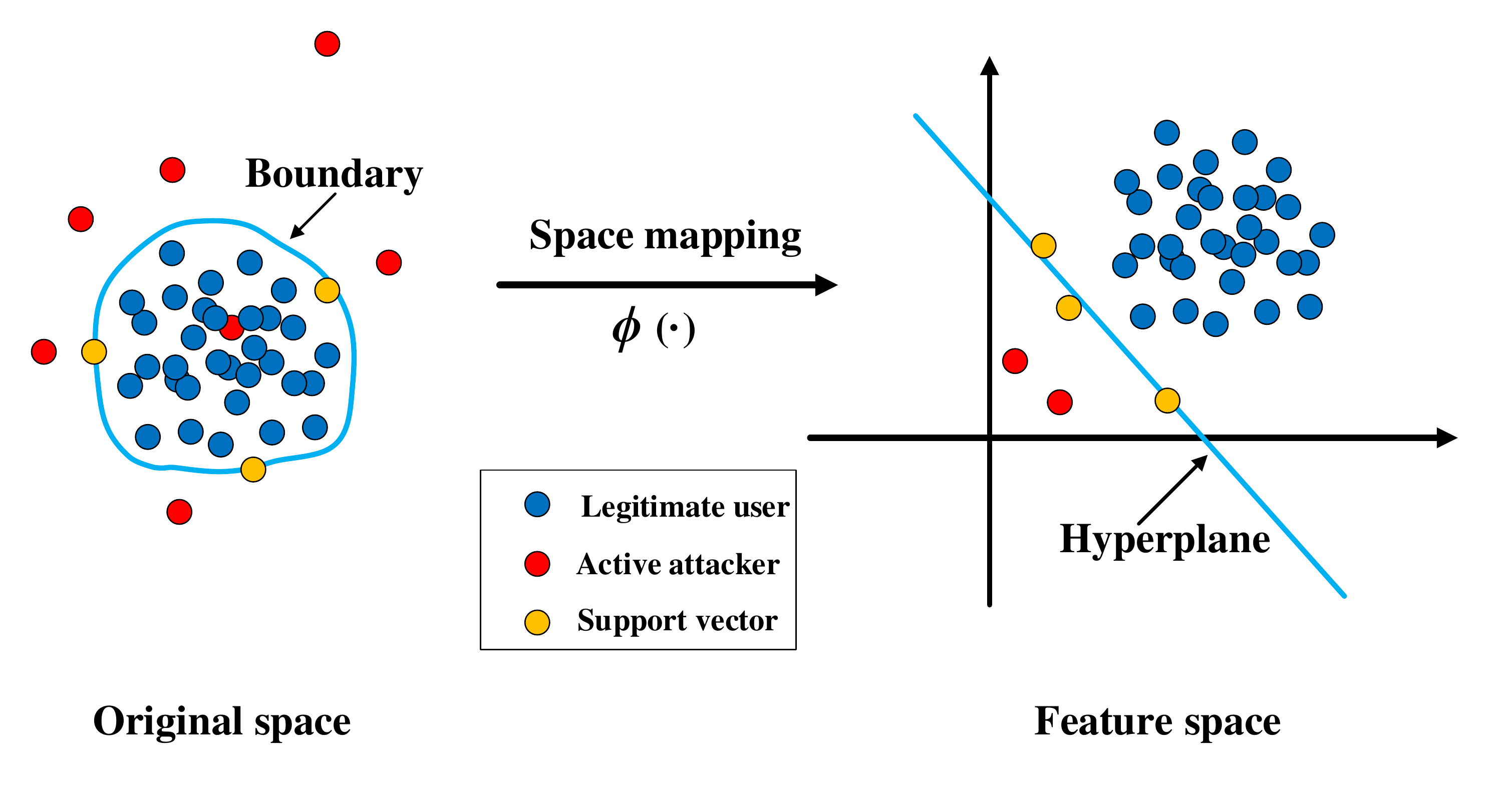} %\vspace{-0.2cm}
	\caption{The profiles of the legitimate user can be mapped into a feature space and separated from the attacking samples.}
	\label{fig:SVM}
	%\vspace{-0.3cm}
\end{figure}
In our system, in order to reduce computational complexity, we compute the DTW distances by dividing the feature series into different chunks instead of directly computing using the entire original signal and smoothing signal, ShieldScatter segments the feature series equally into 128 chunks. Then, ShieldScatter computes the DTW distance in each corresponding chunk. As for energy envelope, variance, maximum and minimum, ShieldScatter divides them into 58 chunks and computes the DTW distances, respectively. Accordingly, we can finally obtain a propagation profile with respect to the DTW distance, which is a vector with the size of 488. Compared with the method of calculating correlation directly, DTW mitigates the effects caused by misalignment.

\subsection{One-Class SVM Classification}\label{sec:SVM}
Based on the similarity comparison of the extracted features, we obtain a propagation profile vector with the size of 488 for every processing. As mentioned, the signals from same devices will experience the same multipath caused by the intentional deployment tags, which will lead to high similarity and short DTW distance for the features. Otherwise, the DTW distances will be significant difference. Thus, we can transfer our problem of detecting the suspicious signals into the problem of distinguishing the propagation profile vectors so as to defend against the attacking signals.

At an intuitional level, in order to distinguish the propagation profile vectors, a most likely method is to set fixed thresholds for each value in the vector. Then, if all the values are lower than the corresponding thresholds, the signals can be considered as positive samples. However, this method is unreliable, since the received signals will be significantly affected by the environment noise  in dynamic environments, which make it difficult to determine these fixed thresholds.

To distinguish the legitimate user and attacker profiles, ShieldScatter formulates our problem as a one-class classification model. In particular, as illustrated in Figure~\ref{fig:SVM}, given a large number of training profiles as $[\bf{x}_1, \bf{x}_2, ...\bf{x}_\emph{i}... \bf{x}_\emph{l}] $,
where $l\in$ $\mathbb{N} $ is the number of profiles, and $\bf{x}_\emph{i}$ the profile vector of the profile $i$. The size of each propagation profile vector with respect to the DTW distances is 488. As shown in Figure~\ref{fig:SVM}, in the original space~\cite{scholkopf2001estimating}, if the profiles are positive and have short DTW distances, the profiles of these legitimate users will be similar and they will gather together closely. However, the samples that are from the attacker will be away from these positive profiles except for a few outliers that have short DTW distances. Thus, the goal of ShieldScatter is to find an optimal boundary to capture most of the positive samples and able to exclude the negative samples.

In order to define the optimal boundary, the strategy of one-class support vector machine (SVM) is to map the sample from the original space into a feature space using function $\phi(\bf{x})$. As shown in Figure~\ref{fig:SVM}, in the feature space, the samples can be separated by a hyperplane with method of maximum margin, where this hyperplane is defined by some samples called support vectors in the training set. In order to seek out these support vectors, this problem can be formulated as 
\begin{eqnarray}
& \underset{\alpha}{\text{min}} & {{1\over 2 } {\sum_{ij}   {\alpha}_i{\alpha}_j \bf{\emph{k}}(\bf{x}_\emph{i},\bf{x}_\emph{j})  }}, \\
& \text{s.t.} & {0\le{\alpha}_i\le{1\over{vl}   }}, \notag \\
& &  { \sum_{i}{\alpha}_i=1}
\end{eqnarray}
%\begin{equation}
%\\\min_{\alpha} {{1\over 2 } {\sum_{ij}   {\alpha}_i{\alpha}_j \bf{\emph{k}}(\bf{x}_\emph{i},\bf{x}_\emph{j})  }},
%\end{equation}
%subject to,
%\begin{equation}
%\\{0\le{\alpha}_i\le{1\over{vl}   }}, { \sum_{i}{\alpha}_i=1},
%\end{equation}
where $v\in$  (0,1] is an upper bound on the fraction of the outliers and a lower bound on the
fraction of support vectors. $k(.)$ represents the Gaussian kernel, which is defined as 
\begin{equation}
\\\bf{\emph{k}}(\bf{x}_\emph{i},\bf{x}_\emph{j}) ={{\phi}(\bf{x}_\emph{i})}\times{{\phi}(\bf{x}_\emph{j})  } ,  
\end{equation}
\begin{equation}
\\\bf{\emph{k}}(\bf{x}_\emph{i},\bf{x}_\emph{j})=\exp(- {{{|| \bf{x}_\emph{i}-\bf{x}_\emph{j} ||} ^2 }\over{2\sigma^2} }   ).
\end{equation}
Then, the decision function of ShieldScatter is defined as
\begin{equation}
\\f(x)=\text{sgn}({\sum_{i}   {\alpha}_i\bf{\emph{k}}(\bf{x}_\emph{i},\bf{x})  }-\rho),
\end{equation}
where $\bf{x}$ is the sample of the testing profiles, $\bf{x}_\emph{i}$ the $i_{th}$ support vector and $ \rho$ the function bias. Hence, the based on the hyperplane decided by the obtained support vectors, the testing profiles can be classified into either legitimate users or attackers.

\subsection{Security Analysis}\label{sec:security}
We finally integrates ShieldScatter with existing security protocols in the upper layers to enable device authentication and defend against active attacks. Sitting between upper-layer security protocols and PHY signal processing, ShieldScatter conforms to reasoning analogous to existing security protocols but differs in that ShieldScatter takes into account the propagation signatures to secure IoT authentication.

\begin{figure}[t]
	%    \center 
	\subfigure[Deauthentication mitigation.]
	{\includegraphics[width=1.65in]{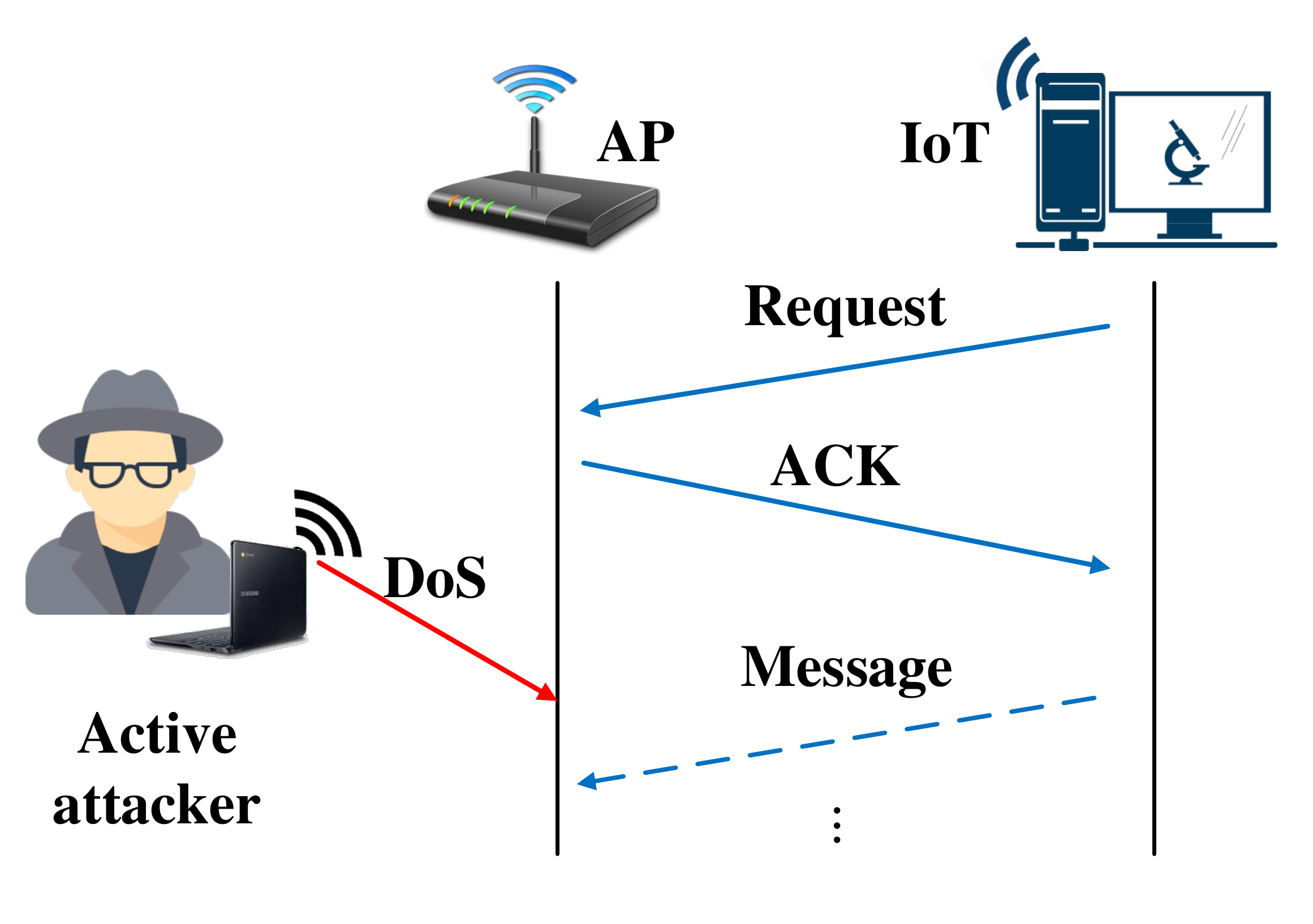}}
	\subfigure[Jamming and replay mitigation.]
	{\includegraphics[width=1.65in]{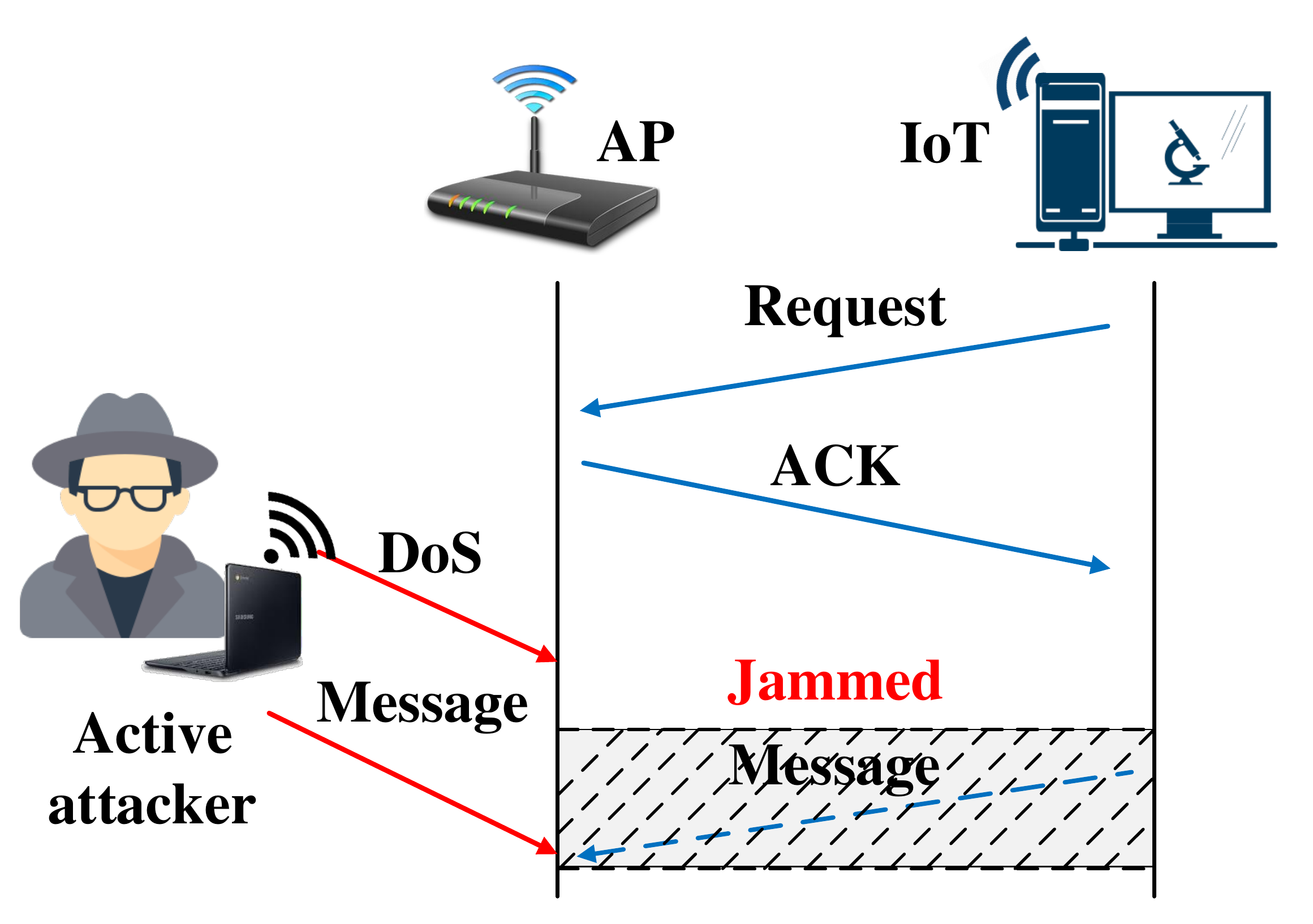}}
	\caption{ShieldScatter integrates with existing security protocols in the upper layers to enable device authentication and defend against active attacks. }
	\label{fig:security}\vspace{-0.3cm}
\end{figure}

\textbf{Deauthentication deadlock mitigation.} There are various ways to launch DoS attacks. A typical type of DoS attacks takes the vulnerability before a secure link has been established. As shown in Figure~\ref{fig:security}(a), we consider that an authentication handshake is in progress. During the authentication handshakes, to deauthenticate the establishment, an attacker can inject an unauthorized deauthentication notification after receiving an acknowledgement (ACK) from the AP, which accordingly leads to a protocol deadlock.

To defend against the deauthentication deadlock, ShieldScatter adds an additional propagation signature processing at the AP with slight protocol changes. Specifically, ShieldScatter controls the message transmission of handshake within the coherence time. Then, upon hearing the deauthentication command, ShieldScatter can compare the similarity between the deauthentication command and the following Message with the operations mentioned in Section~\ref{sec:system}. If the one-class SVM identifies that the deauthentication command is from the attacker, the AP will drop this data frame in the upper layer. Accordingly, ShieldScatter can easily defend against attacks.

\textbf{Jamming and replay mitigation.} An attacker can launch a jamming and replay attack by equipping multiple antennas. A multi-antenna attacker can jam the association packets reception with one directional antenna and records the packet with another antenna. The attacker then replays the recorded packets to the legitimate device. 

As illustrated in Figure~\ref{fig:security}(b), we also take the handshake processing and deauthentication deadlock into account. During this process, an attacker first injects an unauthenticated deauthentication notification after receiving the ACK from the AP. When detecting the following Message, the attacker jams this reception at the AP with one directional antenna, while at the same time records Message with another directional antenna. The attacker then replays the recorded Message to the AP. Thus, both of deauthentication command and Message are from the attacker and the multi-path signatures will be the same. However, when the attacker jams the reception of Message, the multi-path signatures at each tag during jamming are the superposition of the legitimate user and attacker. This will lead to a large difference in the energy for each tag. Thus, ShieldScatter can easily detect this difference and defend against the attacks.   
  \begin{figure}[t]
 	\center
 	\includegraphics[width=2.5in]{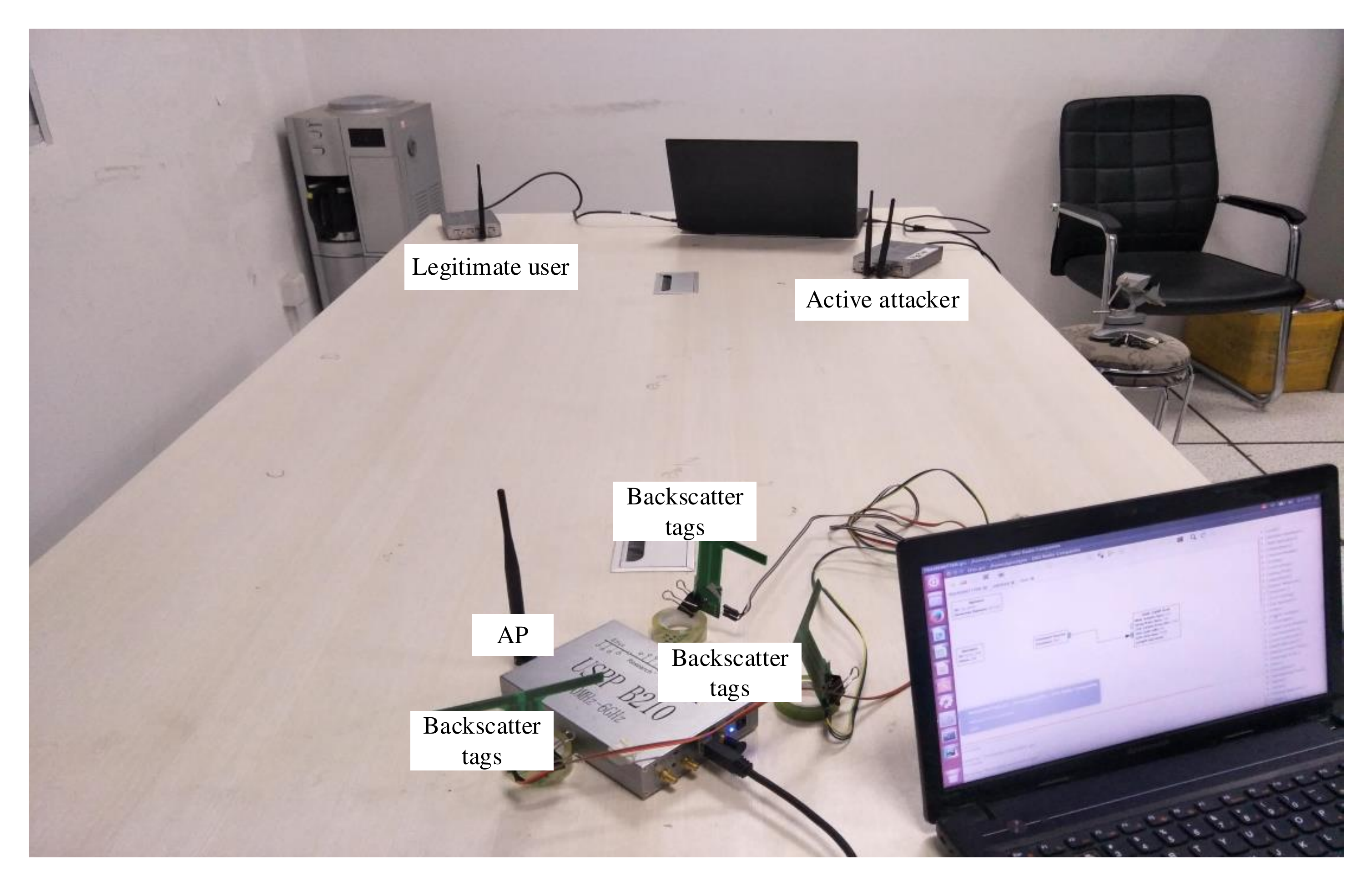} %\vspace{-0.2cm}
 	\caption{We employ two USRPs to act as the AP and legitimate user at a distance of 2.5~m. At the same time, several tags are deployed around the AP to create multi-path signatures. Besides, another two-antennas USRP is used to act as active attacker. }
 	\label{fig:imple}\vspace{-0.3cm}
 \end{figure}
\textbf{Channel spoofing mitigation.} Using wireless physical layer information for location distinction has been explored for many years~\cite{patwari2007robust,li2006securing, liu2010authenticating}. It has been discovered that the characteristics of the wireless channel will be uncorrelated every a half carrier wavelength over distance~\cite{he2013link}. However, the work in~\cite{fang2014you} has found new attacks against these approaches by emulating the multi-path signatures. 

The advantage of Shieldscatter is that this method make it difficult for the attackers to select the real multi-path signatures created by the backscatter tags. Specifically, in our system, ShieldScatter can randomly control the order of tags, as long as it guarantees that the order of the tags is the same in every process. Even though the attacker has emulated all the multi-path signatures, it still cannot decide which multi-path signature is the true one in each time when attacking. Thus, ShieldScatter is more reliable than the methods that simply compare the channel correlation.

\section{Implementation}\label{imple}
As shown in Figure~\ref{fig:imple}, the prototype of ShieldScatter is implemented using multiple backscatter tags and three GNURadio/USRP B210 nodes. The backscatter tags are implemented according to~\cite{liu2013ambient}. We tailor the antenna design to allow tags to work at 900 MHz which is a commonly used frequency for IoT devices. All the tags are deployed around the AP at a distance of 15~cm (i.e., half of the wavelength). In our experiment, both tags transmit data with a bitrate of 10 kbps. Besides, one USRP node equipped with two antennas acts as an active attacker, who monitors RSS variations with one of the antennas while transmitting fake data using the other antenna. The other two USRP nodes are used as the legitimate user and AP, respectively and each of them contains only one antenna. In our experiment, we require the legitimate user and AP to complete the challenge-response protocol in 100~ms so as to maintain the channel to be stable.
\begin{figure}[t]
	\center
	\includegraphics[width=3.2in]{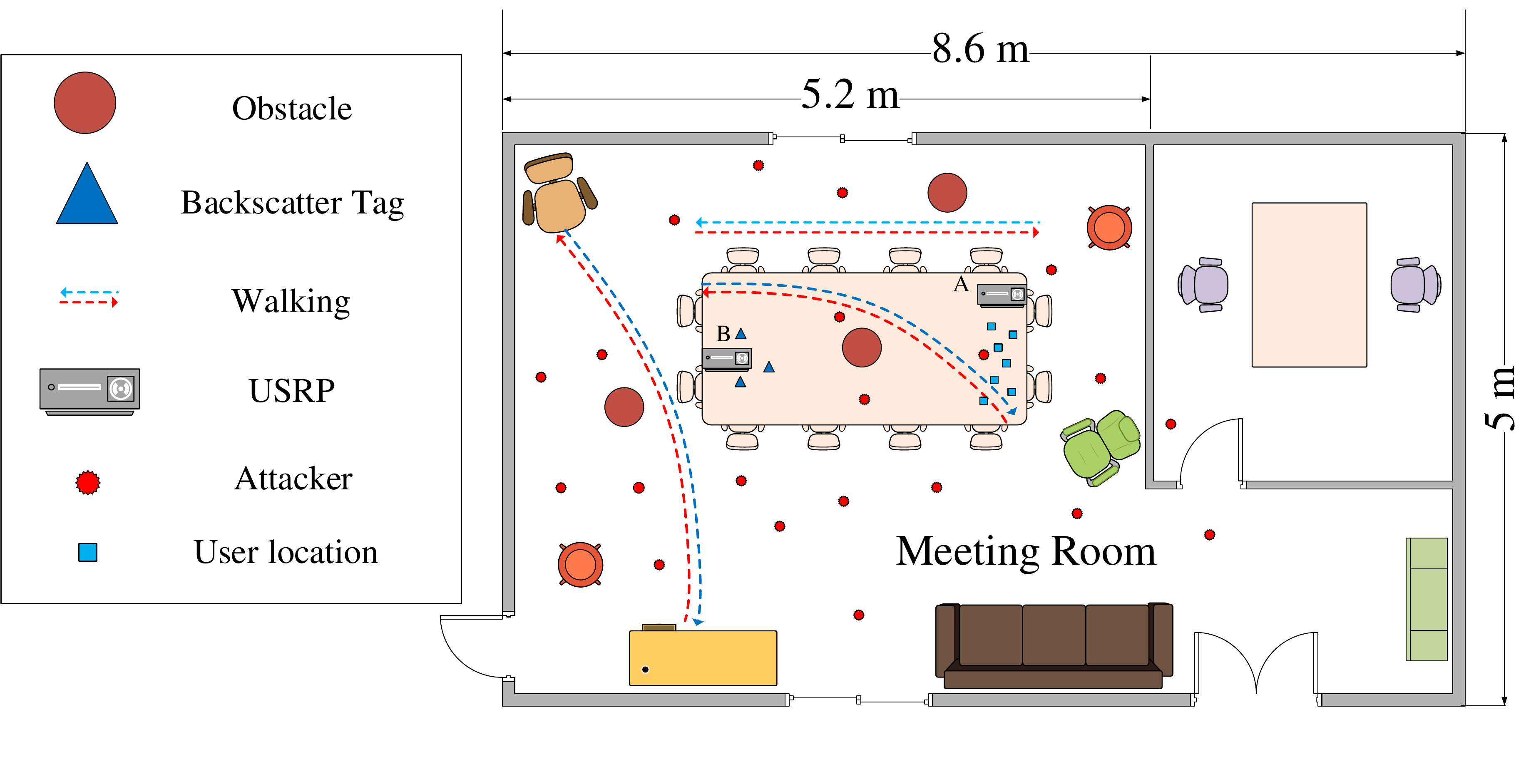} %\vspace{-0.2cm}
	\caption{Floor plan of our evaluation environment. A and B represent the places of legitimate user and AP, respectively. }
	\label{fig:floorplan}\vspace{-0.3cm}
\end{figure}
\section{Evaluation}\label{sec:eva}
In our experiment, we evaluate the performance of ShieldScatter in both static and dynamic environments as shown in Figure~\ref{fig:floorplan}. We employ two USPRs to emulate a legitimate user and an AP which are deployed at a distance of 2.5~m. Specifically, we place the user at different locations (e.g., the blue blocks)  and the AP at location B, respectively. Besides, another USRP that contains two antennas is deployed at different locations (e.g., the red dot in Figure~\ref{fig:floorplan}) to act as an active attacker. In static environment, we conduct our experiment and collect the signals during the day and at night. Besides, we also evaluate the performance in common home environments where we consider the legitimate user and AP are in both line-of-sight (LoS) and none line-of-sight (NLoS) scenarios. In order to construct a NLoS environment, we deploy different kinds of obstacles between the legitimate users and AP. As for dynamic environment, two people are asked to walk around when we perform the experiments.

\begin{figure}[t]
	\center
	\includegraphics[width=2.6in]{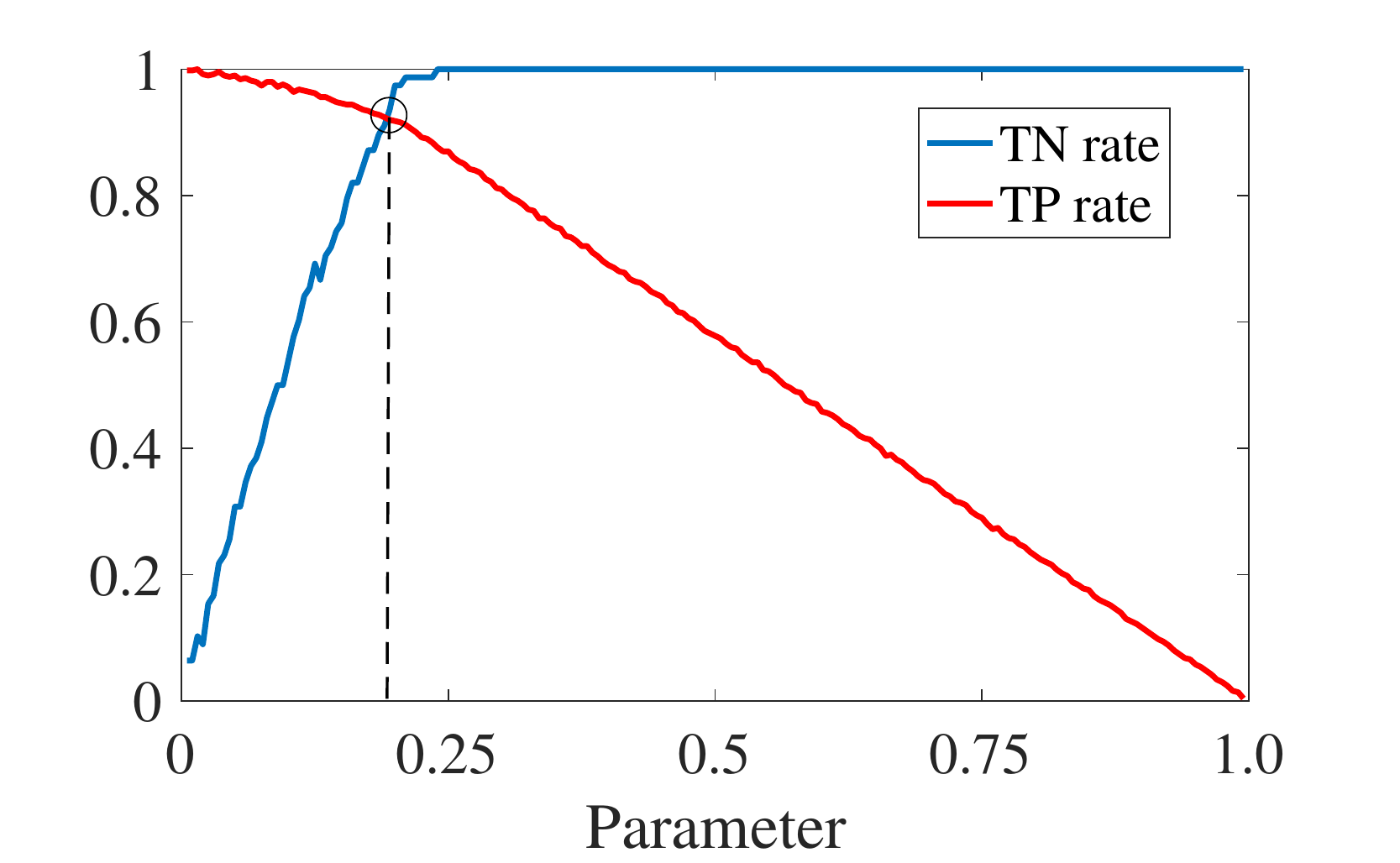} %\vspace{-0.2cm}
	\caption{The performance with respect to the varying parameter $v$. }
	\label{fig:parameter}\vspace{-0.3cm}
\end{figure}
In our experiment, a total number of 1,700 propagation signatures are collected within a month for ShieldScatter in both static and dynamic environments. Then, ShieldScatter extracts the features to construct the profiles for all of these data samples. 577 of the propagation profiles are used to train and construct our one-class SVM model and the rest of the profiles are used to test the performance of ShieldScatter.  

\textbf{Metrics.} We employ the following metrics to evaluate the performance of our system. \begin{itemize}
	\item \textbf{True positive rate.} True positive (TP) rate is defined to be the ratio of the number of propagation profiles in which the samples from the legitimate user is correctly detected to the total number of samples.
	\item \textbf{False positive rate.} False positive (FP) rate is the ratio of the number of propagation profiles in which the samples from the active attackers is falsely recognized as being the legitimate user to the total number of samples.
\end{itemize}

\subsection{Parameter Determination}\label{sec:parameter}
The first step in our experiment is to determine the parameter $v$ for the one-class SVM model. As mentioned before, $v$ is a significant parameter to constrain the bound of outliers and support vectors, and the range of $v$ is $v \in (0,1]$. In order to determine the parameter $v$, we exploit 500 groups of positive data samples and 77 groups of negative samples for input to training the one-class model ranging from 0 to 1 for the parameter $v$. As shown in Figure~\ref{fig:parameter}, it is obvious that the accuracy of correctly detecting the legitimate user decreases with the parameter $v$. However, the accuracy of detecting the active attackers increases when the parameter $v$ grows. That is because when parameter $v$ increases, the bound between the legitimate samples and attacker samples tightens. Sequentially, the outliers (i.e., the attacker) are excluded from the positive samples. However, the larger the parameter increases, the smaller the boundary becomes. In that case, some positive samples are recognized as the outliers and excluded from the positive samples. Thus, in order to determine the optimal $v$ for the one-class SVM model, we make a tradeoff between them. As shown in the Figure~\ref{fig:parameter}, we select the parameter $v$ at intersection point between these two curves. Accordingly, we can achieve the accuracy of 93.7\% for legitimate users and active attacker detection when the parameter $v$ is set as 0.16.

\subsection{Static Environment}\label{sec:static}
Based on the determination of parameter $v$, we evaluate the performance of ShieldScatter in the static environment, where we keep the legitimate user, the AP, and the attackers static. Then, we evaluate ShieldScatter with respect to the distance between the legitimate user and active attackers, the effects in both LoS and NLoS scenarios,  and the number of backscatter tags attached around the AP. 

\begin{figure}[t]
	\center
	\includegraphics[width=2.6in]{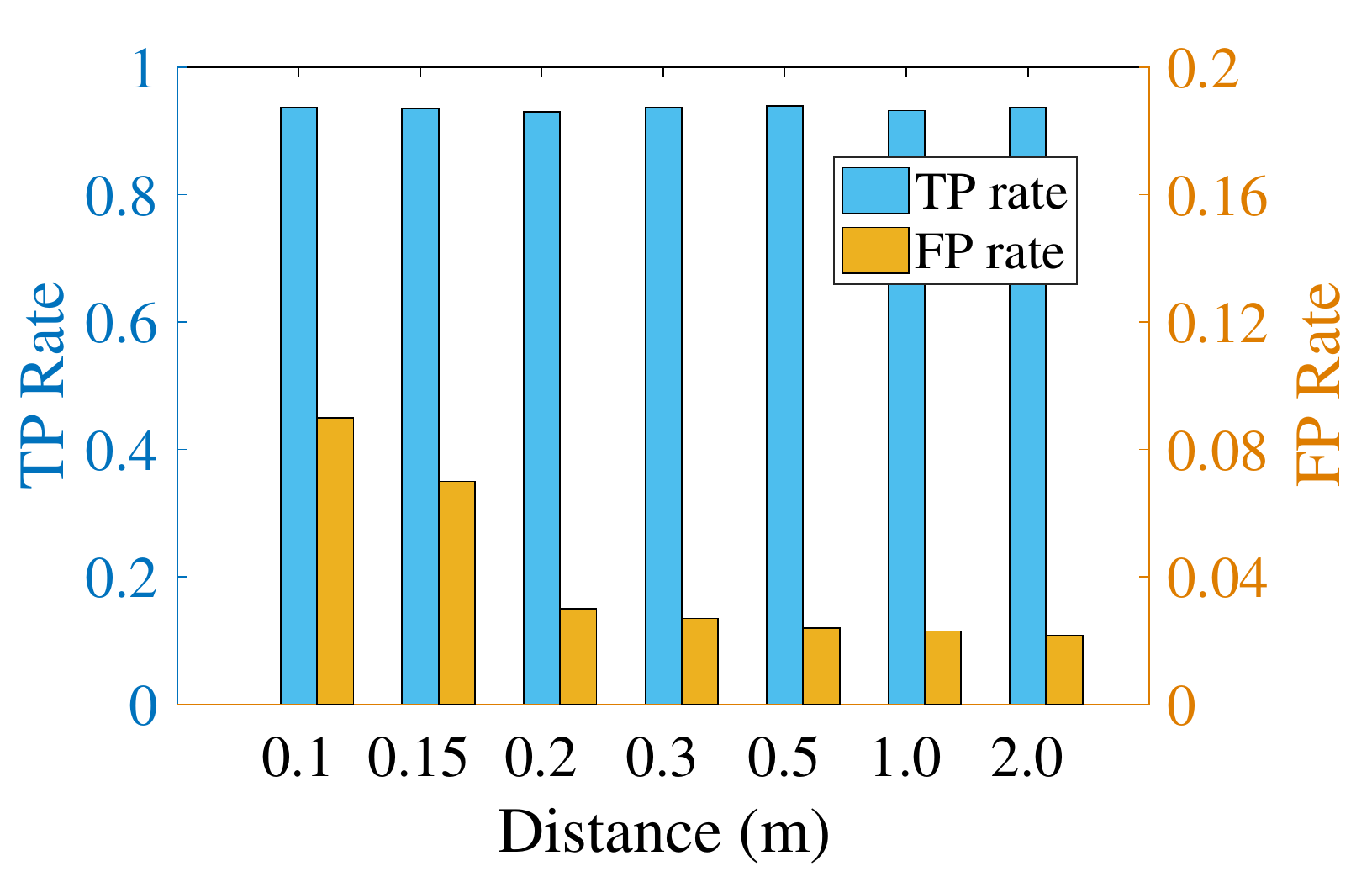} %\vspace{-0.2cm}
	\caption{The performance with respect to the varying distances between the attacker and legitimate user in static environment. }
	\label{fig:static_distance}\vspace{-0.3cm}
\end{figure}

\begin{figure}[t]
	\center
	\includegraphics[width=2.6in]{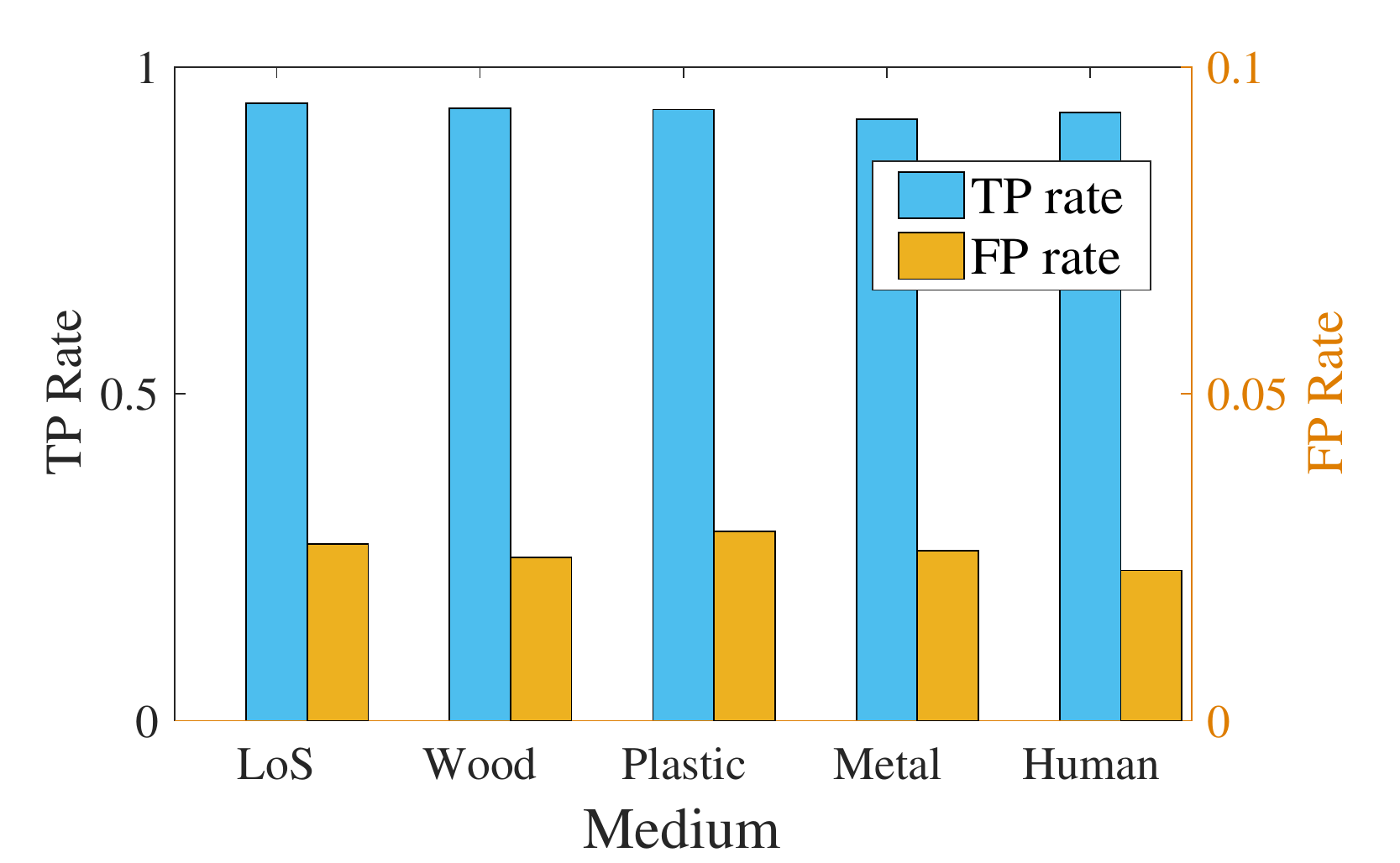} %\vspace{-0.2cm}
	\caption{The performance when the channel is sheltered by different obstacles. }
	\label{fig:medium}\vspace{-0.3cm}
\end{figure}
%\subsubsection{Distance between legitimate user and attackers}\label{sec:distance_static}
\textbf{Distance between legitimate user and attackers.} Usually, the attacker will be far away from the legitimate user. However, if the attacker is small enough and has the ability to get close to legitimate users, it will be a challenge for the IoT devices. Thus, in order to defend against the attackers that are close to the legitimate users, we first evaluate the performance of our system combined with the different distance between the legitimate users and active attackers. In particular, we evaluate the performance in different distances and different directions between the legitimate users and AP ranging from 10~cm to 2~m. After that, we collect the data to construct the profiles as input to test our system.

As shown in Figure~\ref{fig:static_distance}, it is obvious that when the attacker is far away from the legitimate user, ShieldScatter can achieve an average TP rate of 93.6\% and FP rate of 3\%. However, if the legitimate user is close to the attacker, especially when the distances are lower than 15~cm, the FP rate increases dramatically. That is because if the attacker is close enough to the legitimate user, the multi-paths of them caused by the backscatter tags are extremely similar. However, in our experiment, even when the distance is closer to 15~cm, we can still achieve an average FP rate lower than 10\%, which is acceptable for the daily smart home loT devices.

\begin{figure}[t]
	\center
	\includegraphics[width=2.6in]{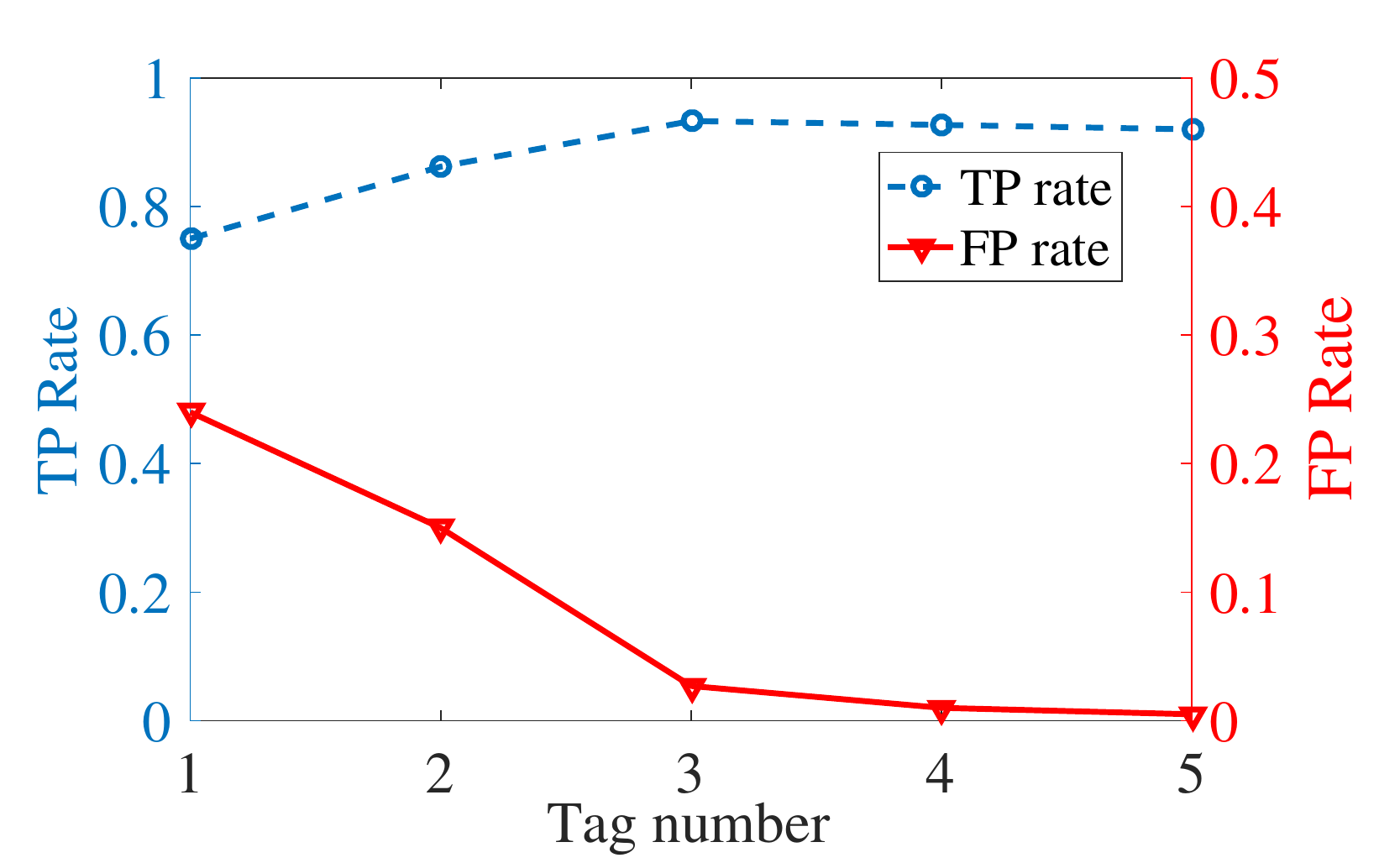} %\vspace{-0.2cm}
	\caption{The performance with respect to the varying number of backscatter tags. }
	\label{fig:tags_num}\vspace{-0.3cm}
\end{figure}
%\subsubsection{ Effects of LoS and NLoS }\label{sec:LoS}
\textbf{Effects of LoS and NLoS scenarios.} For smart home devices, a common situation is that the direct path will be sheltered in some cases. For example, different obstacles will shield the path between the legitimate user and the AP, which accordingly has significant impacts on the backscatter signals. Thus, in order to overcome this predicament, we next evaluate the performance of ShieldScatter in both line-of-sight (LoS) and non-line-of-sight (NLoS) scenarios. Specifically, in order to emulate situations of LoS and NLoS, we exploit the different kinds of common obstacles in daily life, such as wood, plastic, metal and the human body, to shield the direct path between the legitimate user and the AP. Then, we test the performance of our system.

As shown in Figure~\ref{fig:medium}, we can yield an average TP rate of 93.65\% and FP rate lower than 3.1\% in the LoS scenarios. However, if there are obstacles shielding the path between them, the accuracy of legitimate user detection slightly decreases, especially for the metal medium. That is because the metal, a conducting medium, has a shielding effect on the radio propagation. However, we can still achieve an average TP rate of 92\% and FP rate lower than 5\%, which is acceptable and we can confirm that ShieldScatter is reliable even though the channel is sheltered by the daily obstacles. 

%\subsubsection{ Number of backscatter tags}\label{sec:tags}
\textbf{Number of backscatter tags.}
In our experiment, the number of the backscatter tags used to deploy around the AP and construct multi-path signatures is an important factor for the radio propagation. Thus, we then evaluate the performance when using a different number of backscatter tags attached around the AP. 

As shown in Figure~\ref{fig:tags_num}, we can observe that if we just attach one or two backscatter tags on the AP, ShieldScatter can achieve an average FP rate higher than 10\%. It is because the active attacker can easily calculate the distance and power between the legitimate user and the AP. Then the attacker can carefully select the transmitting power and locations for attacking. However, if we deploy three tags, the attacker is harder to keep the arrived power for each tag being similar to the power from the legitimate user. Hence, we can achieve an average TP rate as high as 93.7\% and FP rate lower than 3\%. In addition, if we exploit too many tags (e.g., 4 tags), it leads to strict constraints to the received signal, and accordingly achieve a lower FP rate and lower TP rate. Therefore, three backscatter tags are a appropriate choice in our system.
	
\begin{figure}[t]
	\center
	\includegraphics[width=2.6in]{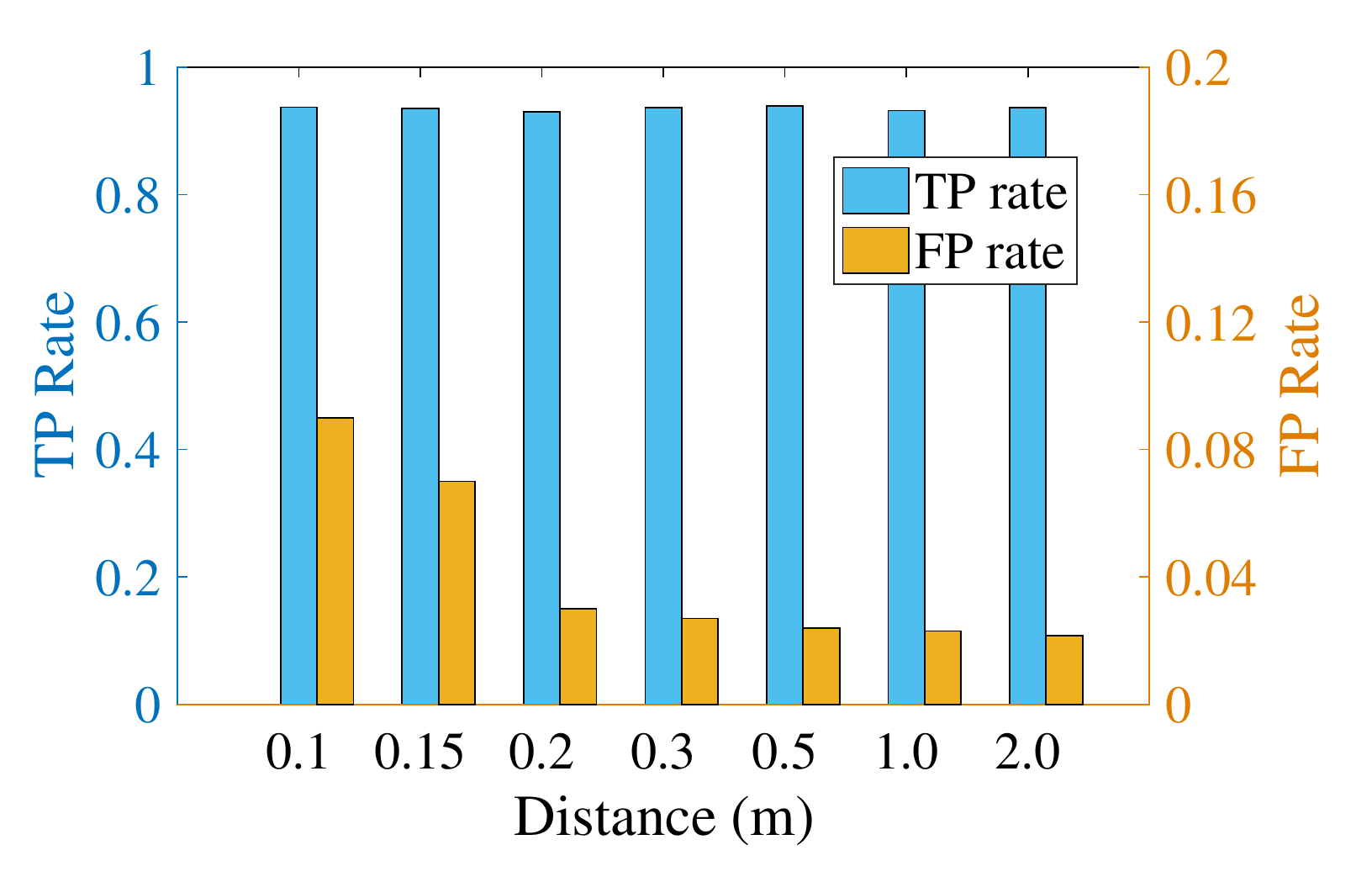}\vspace{-0.2cm}
	\caption{The performance with respect to the varying distances between the attacker and legitimate user in dynamic environment. }
	\label{fig:mobile_distance}\vspace{-0.3cm}
\end{figure}
	
	\subsection{ Dynamic Environment }\label{sec:mobile}
	\textbf{People walking around.} ShieldScatter considers another practical environment for the daily smart home devices, that is, the scenario where some people are walking around.  As shown in Figure~\ref{fig:floorplan}, in order to emulate the dynamic environment, two volunteers are asked to walk around the devices, approach the user, go across the channel and carry out the daily activities. Then,  we evaluate the performance with respect to different distances between the legitimate user and attacker.
	
	As shown in Figure~\ref{fig:mobile_distance}, compared with the results in static environment, when the environment is dynamic, the TP and FP rates of ShieldScatter have slight fluctuation caused by environment noise. However, when we exploit the filter to remove the noise caused by the dynamic effects, ShieldScatter can still maintain an average TP rate higher than 91\%, which is acceptable for the smart home IoT devices. Besides, ShieldScatetr achieves an average FP rate lower than 1.9\%, when the distance is larger than 20 cm. That is because the channel fluctuation makes it more difficult for the attacker to emulate the power for each tag. Thus, our system can remain reliable even in the dynamic environment.
	
	\textbf{Slight movement of the legitimate user.} ShieldScatter takes into account the case that the smart home devices are slightly moved. Specifically, as shown in Figure~\ref{fig:floorplan}, the legitimate user is first placed at location A, and then it is moved to a different place as the blue blocks. Then, we evaluate the performance of our system with respect to the distance between the location A and the user.
	
\begin{figure}[t]
		\center
		\includegraphics[width=2.6in]{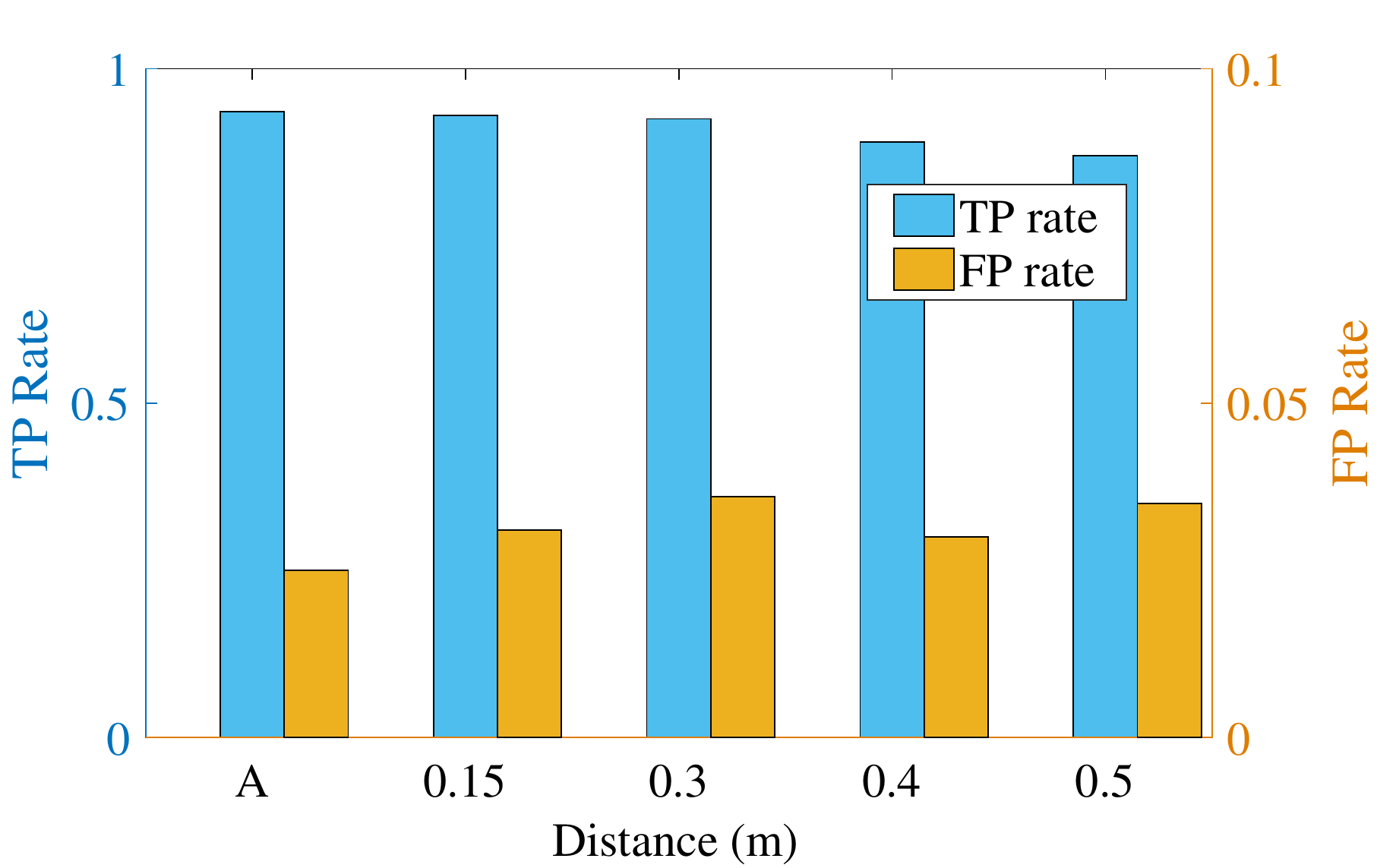} %\vspace{-0.2cm}
		\caption{The performance when the legitimate user is slightly moved. }
		\label{fig:movement}\vspace{-0.3cm}
	\end{figure}

	As shown in Figure~\ref{fig:movement}, we can achieve an average FP rate lower than 3\% when the user is moved to different locations. On the other hand,  when the legitimate user is placed at location A, we can achieve a TP rate of 93\%. Then, if we move the user within a short distance (e.g., within 30 cm), the TP rate remains stable. When the user moves to a longer distance, the TP rate would decrease. However, we can still achieve a TP rate larger than 87\% even though the movement distance is 50 cm. Therefore, slight movements of the legitimate user are allowed in our system.
	
	\begin{figure}[t]
		\center
		\includegraphics[width=2.6in]{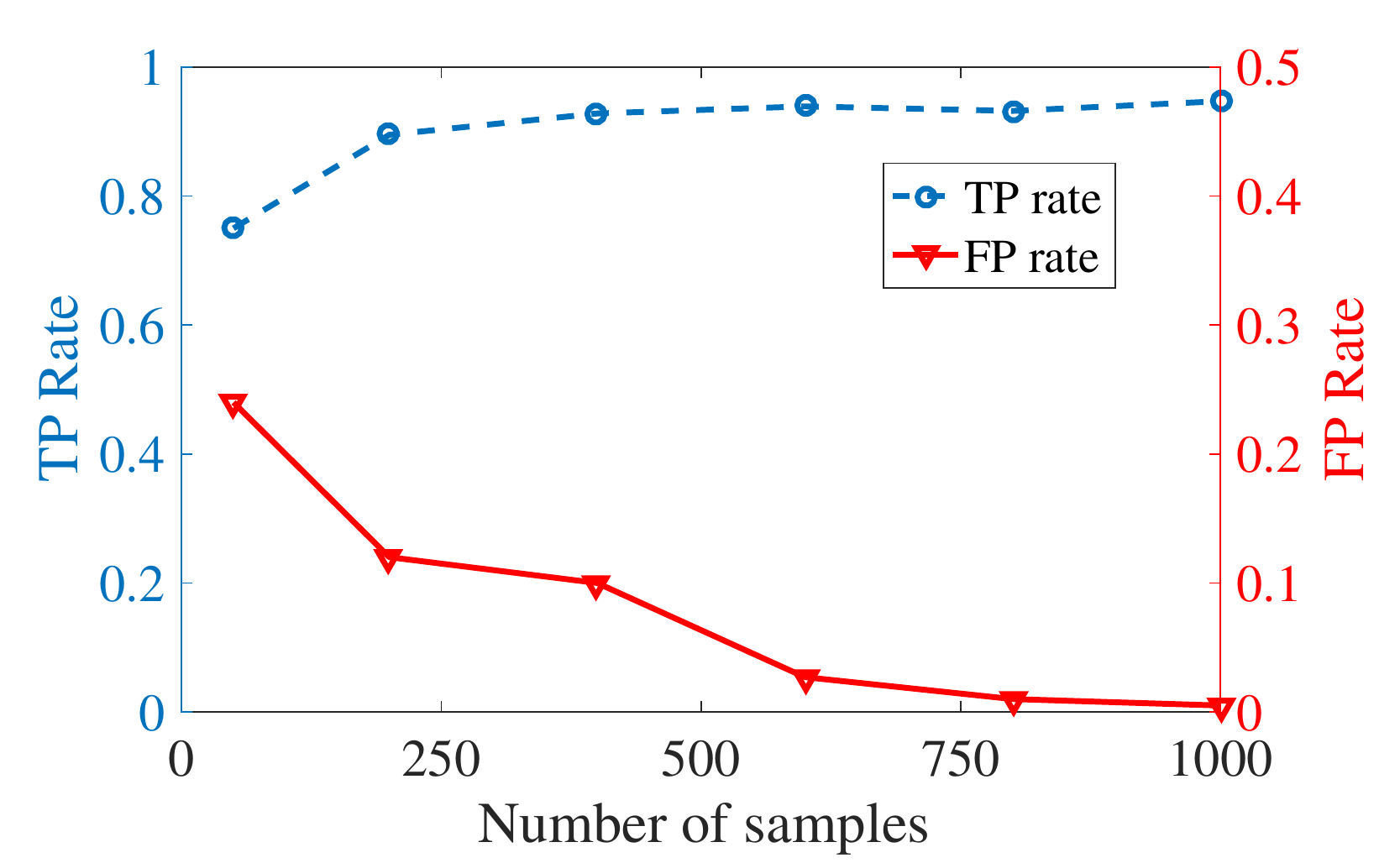} %\vspace{-0.2cm}
		\caption{The performance with respect to the varying number of data samples to train the model. }
		\label{fig:samples_num}\vspace{-0.3cm}
	\end{figure}
		
		\subsection{ Impact Factors for SVM }\label{sec:impact}
		In this section, we evaluate the following two key impact factors on our one-class SVM system.
		
		%\subsubsection{ training size }\label{sec:size}
		\textbf{Training size.}
		In our experiment, ShieldScatter needs to exploit a training set to train a one-class SVM classifier. Then, based on well-trained classifier, we test the testing profiles to detect the legitimate user and defend against the suspicious signals. Thus, it is necessary to select appropriate number of the training set to train and construct the classifier for ShieldScatter. In particular, we train our model with respect to different number of profiles ranging from 50 to 1,000 samples.
		
		As described in Figure~\ref{fig:samples_num}, when the data samples used to train the one-class SVM model are less than 200, ShieldScatter can achieve average TP rate lower than 90\%. However, when we adopt the training samples larger than 600, we can achieve a relatively reliable TP rate of 93.6\% and FP rate lower than 3\%. Consequently, we leverage 577 samples to train and construct our profile.

		%\subsubsection{ The ratio of positive to negative samples}\label{sec:LoS}
		\textbf{The ratio of positive to negative samples.}
		Based on the priori knowledge about one-class SVM, the ratio of positive to negative samples that used to train the model is an important factor. Thus, we evaluate the performance combined with different ratio of between them. Additionally, a noticeable constraint for one-class SVM model that the ratio of positive to negative samples should be very large. In other words, one-class SVM model generally makes use of large number of positive and a few or even no negative samples to train the model. Thus, we study the performance of ShieldScatter with respect to the low ratio of positive to negative samples ranging from 0.05 to 0.5.
		
\begin{figure}[t]
	\center
	\includegraphics[width=2.6in]{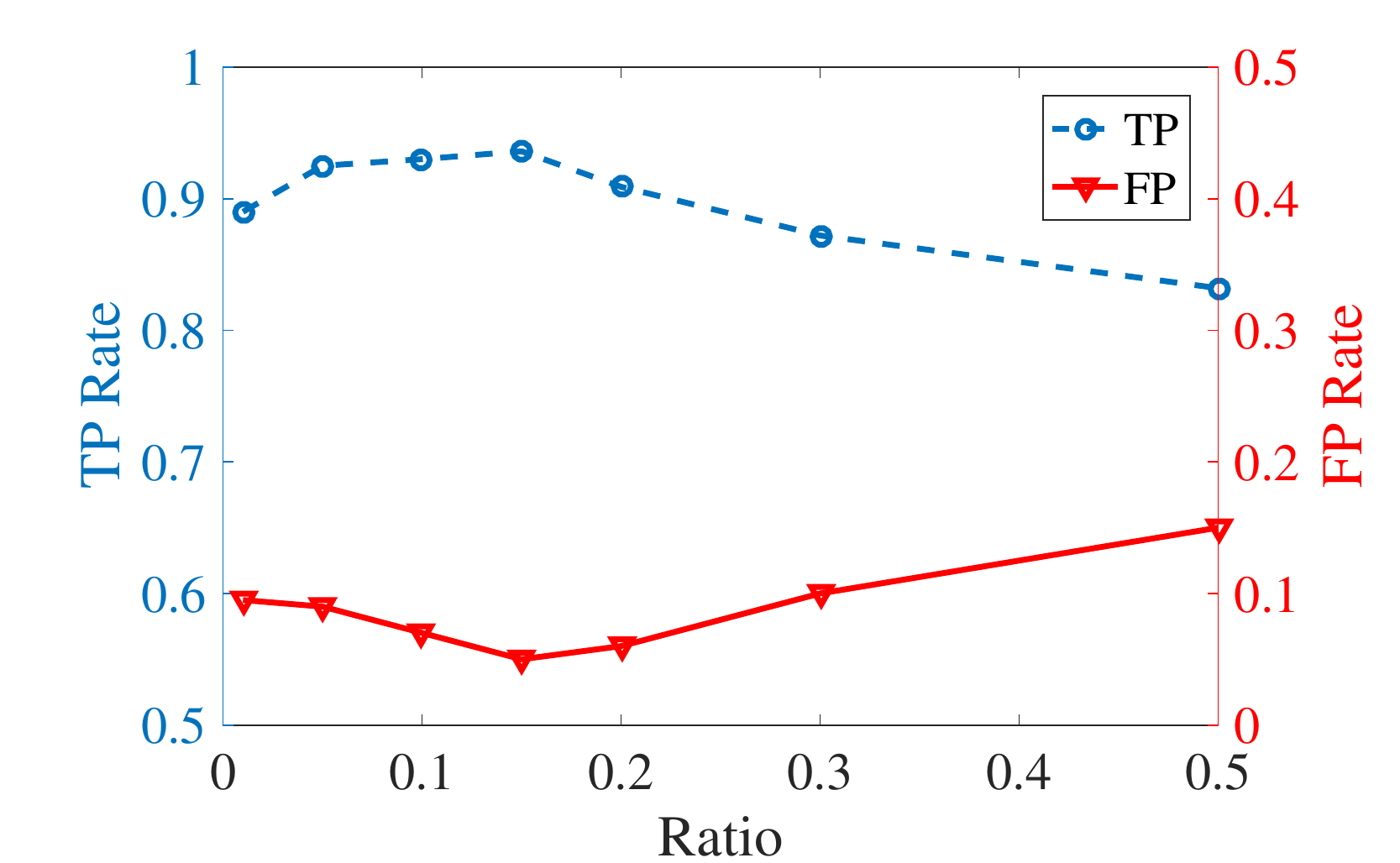} %\vspace{-0.2cm}
	\caption{The performance with respect to the varying ratios of positive to negative samples.}
	\label{fig:ratio}\vspace{-0.3cm}
\end{figure}

		As presented in Figure~\ref{fig:ratio}, when the number of the negative samples is too small, for example, the ratio of positive to negative samples is lower than 0.05, ShieldScatter achieves an average TP rate lower than 90\% and FP rate larger than 10\%. This is because if the negative sample are too small, the suspicious signals on the bound are circled in legitimate user but the positive samples are moved out. Conversely, if the number of the negative samples are too large, the model is unable to distinguish the positive and negative samples, which will lead to lower detection accuracy for the model. Therefore, ShieldScatter selects the ratio of positive to negative samples as low as 0.154 for the model training.			
	
\section{Related work}\label{sec:related}
\textbf{ Backscatter communications.}
Backscatter communication has been considered as a promising communication mechanism in the future. Ambient backscatter originates from the RFID systems that makes use of RFID readers to provide power and communicate with battery-free tags~\cite{wang2012efficient}. The difference between them is that backscatter can harvest ambient RF signal and enable two RF-powered devices to communicate by scattering and creating the path on the ambient signals ~\cite{liu2013ambient}. Besides, in order to enable different RF-powered devices to communicate with each other, WiFi backscatter, FM backscatter and FS backscatter are proposed~\cite{kellogg2014wi,wang2017fm,zhang2016enabling}, which makes backscatter communication become applicable for current IoT devices. In our study, ShieldScatter employs the backscatter tags as~\cite{liu2013ambient} to create significant multi-path propagation signatures and construct unique profile for the IoT device.

\textbf{Physical-layer propagation signatures.}
Recent years, fine-grained physical-layer propagation signatures ~\cite{liu2012enhanced,wang2018securing,xiao2009channel,zhang2008advancing}have been successfully used to secure wireless system. For example, SecureArray~\cite{ xiong2013securearray} secures WiFi by using AoA information to construct sensitive signatures. Besides, some researchers seek for other signatures, such as received signal strength (RSS) for user authentication~\cite{cai2011good,chandrasekaran2009detecting}. Proximate~\cite{mathur2011proximate} securely pairs two devices in proximity within a half-wavelength distance by comparing their RSS variations. Wanda~\cite{pierson2016wanda} employs two antennas to authenticate the devices in proximity according to the large RSS variations between the two antennas. However, these studies need two or more antennas to construct the signatures, which will be not appropriate for the smart home IoT devices that contain only one antenna. Different from the past works, ShieldScatter secures the IoT devices without additional antennas and it can still achieve high detecting accuracy with several low-cost backscatter tags for assistance. 

\textbf{Wireless localization.}
Accurate localization can help detect the signals and secure the IoT devices. SpotFi~\cite{kotaru2015spotfi} can achieve high localization accuracy for the devices by combining AoA signatures with time of flight (ToF). WiTag~\cite{kotaru2017localizing} localizes the backscatter tags using commodity WiFi signals. Other localization systems, such as RFID, are also explored to localize the IoT devices~\cite{wang2013rf,wang2013dude,yang2014tagoram}. PinIt~\cite{wang2013dude} deploys large number of RFID tags around the devices and then exploits multipath to localize the target that has the similar multipath profiles. All these RFID-based methods need a dedicated RFID reader to help communicate with the tags.
		
\section{Conclusion}\label{sec:conclusion}			
We present ShieldScatter, a lightweight system to secure IoT devices pairing and data transmission by intentionally creating multi-path signatures by using several low-cost backscatter tags that are attached to an AP or IoT device. ShieldScatter secures IoT devices without using an expensive antenna array or hardware modification to existing devices. We have evaluated the performance of ShieldScatter in both static and dynamic environments. Our results show that even the attacker is located only 15 cm away from the legitimate device, ShieldScatter with merely three backscatter tags can mitigate 97\% of spoofing attack attempts while at the same time triggering false alarms on just 7\% of legitimate traffic.

\begin{acks}
	The research was supported in part by the National Science Foundation of China under Grant 61871441, 61502114,
	91738202, and 61531011, Major Program of National Natural Science Foundation of Hubei in China with Grant
	2016CFA009, Key Laboratory of Dynamic Cognitive System of Electromagnetic Spectrum Space (Nanjing Univ. Aeronaut. Astronaut.), Ministry of Industry and Information Technology, Nanjing, 211106, China with KF20181911.
\end{acks}

%\balance
\bibliographystyle{ACM-Reference-Format}
\bibliography{sample-bibliography}

\end{document}